\documentclass[11pt]{article}
\pdfoutput=1
\usepackage{jheppub}
\usepackage{graphicx}
\usepackage{array,tikz,tikz-cd,float}
\usetikzlibrary{positioning, quotes, decorations.markings, decorations.pathmorphing, shapes, calc}
\usepackage{color}
\usepackage{bm, dsfont, braket, slashed}

\tikzset{middlearrow/.style={
		decoration={markings,
			mark= at position 0.5 with {\arrow{#1}} ,
		},
		postaction={decorate}
	}
}

\newcommand{\pd}{{\partial}}
\newcommand{\opd}{{\overline \partial}}

\newcommand{\wt}{\widetilde}

\newcommand{\diag}{\text{diag}}
\newcommand{\Hom}{\text{Hom}}

\newcommand{\Tr}{\text{Tr}}

\newcommand{\cf}{\emph{cf.}}
\newcommand{\ie}{\emph{i.e.}}
\newcommand{\eg}{\emph{e.g.}}

\newcommand{\be}{\begin{equation}}
	\newcommand{\ee}{\end{equation}}
\newcommand{\bp}{\begin{pmatrix}}
	\newcommand{\ep}{\end{pmatrix}}
\newcommand{\bsp}{\left(\begin{smallmatrix}}
	\newcommand{\esp}{\end{smallmatrix}\right)}

\newcommand*\diff{\mathop{}\!\mathrm{d}}

\newcommand{\D}{{\mathbb D}}
\newcommand{\R}{{\mathbb R}}
\renewcommand{\P}{{\mathbb P}}

\newcommand{\W}{{\mathbb W}}
\newcommand{\C}{{\mathbb C}}
\newcommand{\Z}{{\mathbb Z}}

\newcommand{\CA}{{\mathcal A}}
\newcommand{\CB}{{\mathcal B}}

\newcommand{\CF}{{\mathcal F}}

\newcommand{\CH}{{\mathcal H}}

\newcommand{\CM}{{\mathcal M}}
\newcommand{\CN}{{\mathcal N}}
\newcommand{\CO}{{\mathcal O}}

\newcommand{\CT}{{\mathcal T}}
\newcommand{\CY}{{\mathcal Y}}
\newcommand{\CV}{{\mathcal V}}

\newcommand{\CX}{{\mathcal X}}

\DeclareMathOperator{\Spin}{Spin}

\newcommand{\oz}{{\overline{z}}}

\newcommand{\oQ}{{\overline{Q}}}

\newcommand{\otheta}{{\overline{\theta}}}

\newcommand{\bA}{\mathbf{A}}
\newcommand{\bB}{\mathbf{B}}

\newcommand{\bS}{\mathbf{S}}
\newcommand{\bT}{\mathbf{T}}
\newcommand{\bW}{\mathbf{W}}
\newcommand{\bX}{\mathbf{X}}
\newcommand{\bY}{\mathbf{Y}}
\newcommand{\bGamma}{\mathbf{\Gamma}}

\newcommand{\bPsi}{\mathbf{\Psi}}
\newcommand{\bPhi}{\mathbf{\Phi}}

\newcommand{\bLambda}{\mathbf{\Lambda}}
\newcommand{\bmu}{\boldsymbol{\mu}}
\newcommand{\bnu}{\boldsymbol{\nu}}
\newcommand{\bOmega}{\mathbf{\Omega}}

\newcommand{\Fl}{\textrm{Fl}}

\newcommand{\Sb}{\textrm{Sb}}
\newcommand{\Ff}{\textrm{Ff}}

\newcommand{\dR}{\textrm{dR}}

\newcommand{\Maps}{\textrm{Maps}}
\newcommand{\Sect}{\textrm{Sect}}

\newcommand{\norm}[1]{{{:\!{#1}\!:}}}

\numberwithin{equation}{section}
\numberwithin{figure}{section}
\numberwithin{table}{section}

\title{Vertex Operator Algebras and Topologically Twisted Chern-Simons-Matter Theories}
\author[1]{Niklas Garner}
\affiliation[1]{Department of Physics, University of Washington, Seattle, WA 98195}

\emailAdd{nkgarner@uw.edu}

\abstract{We consider several topologically twisted Chern-Simons-matter theories and propose boundary VOAs whose module categories should model the category of line operators of the 3d bulk. Our main examples come from the topological $A$ and $B$ twists of the exotic $\CN=4$ Chern-Simons-matter theories of Gaiotto-Witten, but we show that there is a topological ``$A$-twist" for a much larger class of $\CN\neq4$ theories. We illustrate a particular example of this new class of theories that admits the $p=2$ singlet VOA $\mathfrak{M}(2)$ on its boundary and comment on its relation to the $\psi \to \infty$ limit of the Gaiotto-Rap{\v c}{\'a}k corner VOA $Y_{1,1,0}[\psi]$.}

\begin{document}
\today
\maketitle

\section{Introduction}

A recurring theme in the study of 3d topological quantum field theories (TQFTs) has been their connection to vertex operator algebras (VOAs). The prototypical example of this phenomenon is the Chern-Simons/Wess-Zumino-Witten (CS/WZW) correspondence \cite{WittenJones, EMSS}, where one finds that many aspects of the bulk Chern-Simons theory are encoded in a boundary WZW model. For example, line operators in the bulk TQFT are realized as modules for the boundary VOA; the quantum state space $\CH(\Sigma)$ of the bulk TQFT on a Riemann surface $\Sigma$ is realized as conformal blocks of the boundary VOA; etc. 

More recently, this avenue has been pursued for 3d TQFTs obtained via topologically twisting supersymmetric theories, in particular twisted 3d $\CN=4$ theories. For such theories, there are two possible topological twists: the $A$ twist, which is a dimensional reduction of the 4d Donaldson-Witten twist \cite{WittenTQFT}, and the $B$ twist, leading to Rozansky-Witten theory \cite{RW} and studied by Blau and Thompson for pure gauge theory \cite{BT}. The topological $B$-twist twists the Lorentz group with the Coulomb-branch $R$-symmetry $SU(2)_B$%
\footnote{The Coulomb branch $R$-symmetry is often denoted $SU(2)_C$. Correspondingly, the $B$-twist is sometimes called the $C$-twist.} %
For 3d $\CN=4$ theories of hypermultiplets coupled to ($\CN=4$) vector multiplets, the corresponding BPS equations require local constancy of the hypermultiplet scalars and forces the gauge fields, complexified by the adjoint scalars in the vector multiplet, to be flat. Similarly, the $A$-twist uses the Higgs-branch $R$-symmetry $SU(2)_A$%
\footnote{As above, the Higgs branch $R$-symmetry is often denoted $SU(2)_H$ and the $A$-twist correspondingly called the $H$-twist.} %
to modify the Lorentz group. The relevant BPS equations impose a Dirac equation on the hypermultiplet scalars, which becomes spinors in the $A$-twist, and allow the gauge field and adjoint scalars to form BPS monopoles.

Much is known about these TQFTs, including their algebras of local operators \cite{BFNII,RW}, categories of line operators \cite{linevortex}, and their state spaces on Riemann surfaces \cite{BFK,BFKHilb}. In \cite{CostelloGaiotto}, Costello and Gaiotto proposed VOAs that play the role filled by the WZW model for the $A$ and $B$ twists of standard 3d $\CN=4$ theories built from hypermultiplets coupled to vector multiplets. The proposed VOAs have been shown to reproduce bulk local operators \cite{CostelloCreutzigGaiotto} and were recently used to better understand the braided-monoidal structure of bulk line operators \cite{BNbraid}. We also note that classical aspects of these boundary conditions, as well as their relation to branes in supergravity and string theory, were recently described in \cite{BLS}.

Unlike the rational VOAs that appear in the study of Chern-Simons theory with compact gauge group, the VOAs that appear in the study of $A$- and $B$-twisted 3d $\CN=4$ theories are nearly always logarithmic. The study of logarithmic VOAs and their relations to 3d TQFTs dates back to the work of Rozansky and Saleur \cite{RStorsion, RSpoly, RSwzw} on $U(1|1)$ Chern-Simons theory, but logarithmic VOAs have appeared in many other intersections of topology and physics, \eg\, \cite{Mikhaylov, MWbranes, 3dmod, 3foldVOA, GaiottoRapcak, GaiottoLanglands, CDGG, GHNPPS}.

The goal of this paper is to extend the work of Costello-Gaiotto to a range of topological Chern-Simons-matter theories. The main class of examples we consider are the $A$ and $B$ twists of the exotic $\CN = 4$ Chern-Simons-matter theories of Gaiotto and Witten \cite{GaiottoWitten-Janus}. The matter multiplets allowed by the Gaiotto-Witten theories are highly constrained; $\CN=4$ supersymmetry requires that the gauge group's action satisfy what Gaiotto-Witten call the ``fundamental identity." Explicitly, the $\CN=4$ Chern-Simons theories of Gaiotto-Witten start with a general $\CN=4$ theory, e.g. an $\CN=4$ theory of hypermultiplets and vector multiplets, with an $H_c$ Higgs (or Coulomb) $\CN=4$ flavor symmetry. As $H_c$ is an $\CN=4$ flavor symmetry, there are a triplet of real scalars $\nu_i$ -- the hyperk\"ahler moment maps for the $H_c$ symmetry -- and the existence of $\CN=4$ supersymmetry after gauging with Chern-Simons gauge fields requires that the corresponding complex scalar $\nu = \nu_1 + i \nu_2$ satisfies
\be
	K(\nu,\nu) = \textrm{ constant}
\ee
and similarly for any other choice of complex structure on the Higgs (or Coulomb) branch, where $K$ is the level of the Chern-Simons gauge fields; this is the fundamental identity of \cite{GaiottoWitten-Janus}. The constant on the right-hand side can depend on the choice of complex structure and vanishes many examples of interest. For hypermultiplets in a linear representation $R$, the fundamental identity requires $R$ be the odd part of an extension of the (complexified) gauge group $H$ to a Lie supergroup $\widehat{H}$ and are classified by (possibly products of) the supergroups $GL(N|M)$ and $OSp(N|2M)$.

Many other aspects of these field theories have been studied, including their BPS Wilson lines \cite{KapustinSaulina-CSRW, KWY, DTThyperloops, OWZ, CDTprofusion, MOPWZ, roadmap}, and in some instances their partition functions, superconformal indices, and moduli spaces \cite{ABCD, NosakaYokoyama1, NosakaYokoyama2,ImamuraKimura, JafferisYin}. Kapustin and Saulina initiated the study of the $B$ twist of the Gaiotto-Witten theories in \cite{KapustinSaulina-CSRW}, calling them Chern-Simons-Rozansky-Witten theories, being an amalgam of the two topological theories. For the class of theories labeled by a Lie supergroup $\widehat{H}$, Kapustin-Saulina identify the $B$-twisted theory with a partially gauge-fixed Chern-Simons theory for the supergroup $\widehat{H}$. Work of K\"{a}ll\'{e}n-Qiu-Zabzine \cite{KQZ} realized these $B$-twisted theories as instances of the AKSZ construction \cite{AKSZ} and discussed aspects of the corresponding 3-manifold invariants.

In this paper, we find an unexpected family of topological deformations to theories that \emph{do not} come from an underlying $\CN=4$ theory. In Section \ref{sec:CSYMA} we gauge a Higgs-branch flavor symmetry of a general $A$-twisted $\CN=4$ Yang-Mills gauge theory with Chern-Simons gauge fields, without regard for the fundamental identity. This is not expected to be an $\CN=4$ theory in general, or even flow to one in the IR. Nonetheless, we find that the resulting theory is topological independent of the fundamental identity.%
\footnote{Somewhat more precisely, we have only checked that these theories are perturbatively topological, \cf\, \cite[Section 2.3]{twistedN=4}; we are optimistic that this persists for the full, non-perturbative theory. As described in \cite[Section 2.3.2]{CostelloDimofteGaiotto-boundary}, the topological nature of the 3d bulk implies that any boundary VOA necessarily has a holomorphic stress tensor. So long as the boundary VOAs described in \ref{sec:bdyCSYMA} are concentrated in cohomological degree 0, they should admit a description as a coset and therefore have natural holomorphic stress tensors -- the coset stress tensor. The existence of a holomorphic stress tensor in these boundary VOAs is a strong indication that the bulk is indeed topological.} %
In particular, we expect to find 3d TQFTs given by gauging (perhaps a subgroup of) the Higgs-branch flavor symmetry of \emph{any} $A$-twisted $\CN=4$ super Yang-Mills theory with Chern-Simons fields.

This enormous class of TQFTs does not seem to appear in the literature in this generality, although certain examples briefly appear in \cite{KLLtop}, and we begin the analysis of some important aspects thereof; namely, identifying a tractable boundary VOA whose module category should capture the category of line operators in the bulk TQFT. It is worth noting that in special cases the underlying supersymmetric gauge theories, $\CN = 2$ Chern-Simons fields coupled to hypermultiplets (possibly gauged with $\CN=4$ vector multiplets), recently appeared in \cite{GKLSYrank0} and were used to extract modular data of certain semisimple but non-unitary 3d TQFTs. We consider their example of $U(1)_1$ coupled to a single hypermultiplet (which is expected to flow to an $\CN=5$ SCFT \cite{GLMMS}) in Section \ref{sec:gaugedhyper}, finding that it admits Kausch's $p=2$ singlet VOA $\mathfrak{M}(2)$ as a boundary VOA. We also briefly describe how this example arises as the $\psi\to\infty$ limit of a corner construction \cite{GaiottoRapcak} in 4d $\CN=4$ $U(1)$ gauge theory. Since the two TQFTs extracted in \cite{GKLSYrank0} are argued to be the $A$ and $B$ twists of the IR $\CN=5$ theory, it would be interesting to understand the connection between their TQFTs and the one presented in this paper.%
\footnote{It is not immediately clear to us that our ``$A$-twist'' is related to either of the twists described in \cite{GKLSYrank0}. In particular, the supercurrents corresponding their twists have non-trivial monopole number, cf. Eq. (3.27) of \emph{loc. cit.}, whereas our twist is realized by a current with trivial monopole number. Moreover, we find the ``$A$-twist'' independent of the level $k$ of the Chern-Simons fields, whereas their analysis is restricted to $|k| = 1$.} %

In the recent paper \cite{CDGG}, we studied in great detail another collection of examples of the theories considered more generally in Section \ref{sec:CSYMA} of this paper: we used $\CN=2$ Chern-Simons gauge fields to gauge the $SU(n)$ Higgs-branch flavor symmetry of the $\CN=4$ theory $T[SU(n)]$ of \cite{GaiottoWitten-Sduality}. We argued that the $A$ twist of this theory yields a TQFT that is intimately related to those constructed by mathematicians using unrolled quantum groups \cite{BCGPM, BGPMR}, the VOAs arising at corners of $\CN=4$ super Yang-Mills \cite{GaiottoRapcak,CLtriality}, and recent work on Coulomb branches and knot invariants \cite{GHNPPS}. Moreover, we showed that this theory admits the well-known triplet algebras $\mathfrak{W}(p)$ \cite{Kausch} (for $n=2$) and their higher rank analogs, the Feigin-Tipunin algebras $\mathcal{FT}_k(\mathfrak{sl}(n))$ \cite{FTalg}, as boundary VOAs, and conjecture that they are related to the boundary VOAs constructed in Section \ref{sec:CSYMA} via new, logarithmic level-rank dualities.%
\footnote{Really, an orbifold of an extension of this coset VOA is identified with a level-rank dual of the Feigin-Tipunin algebra $\mathcal{FT}_k(\mathfrak{sl}(n))$, although an explanation is beyond the scope of this paper. See \cite[Section 6.5]{CDGG} for more details.} %
The Feigin-Tipunin algebras are themselves conjectured \cite{FGST1,FGST2} to realize a logarithmic version of the Kazhdan-Lusztig correspondence \cite{KLI+II, KLIII, KLIV}, whereby its category of modules (suitably defined) is equivalent to the category of modules the restricted quantum group $\mathfrak{u}_q (\mathfrak{sl}(n))$ for $q = e^{i \pi/k}$ an even root of unity.

\subsection{Outline and summary}
We now briefly outline the structure of the paper and summarize our results.

In Section \ref{sec:twistedVOAs} we review our main tool: the ``twisted superfields" of Aganagic-Costello-McNamara-Vafa \cite{AganagicCostelloMcNamaraVafa}, which we call the ``twisted formalism." This twisted formalism provides a remarkably compact description for the holomorphic-topological ($HT$) twists of a wide range of $\CN=2$ Chern-Simons-matter theories, including those of interest to this paper. The $HT$ twist of a general $\CN=2$ theory was first introduced in \cite{CDFK1, CDFK2} but exists for any $\CN\geq2$ theory, a fact central to our constructions. The twisted formalism of \cite{AganagicCostelloMcNamaraVafa} was recently used by Costello-Dimofte-Gaiotto to describe boundary chiral algebras for a wide range $HT$-twisted $\CN=2$ theories and $\CN=(0,2)$ boundary conditions in \cite{CostelloDimofteGaiotto-boundary}; work of Gwilliam-Williams \cite{GwilliamWilliams} implies that these theories admit a choice of gauge fixing that yields a 1-loop exact, finite perturbative quantization; a recent paper of Zeng \cite{Zeng} used the twisted formalism to describe aspects of the algebra of local operators in $HT$-twisted $\CN=2$ abelian gauge theories; and the companion paper \cite{twistedN=4} applied the twisted formalism to succinctly reproduce aspects of standard $\CN=4$ theories of gauged hypermultiplets.

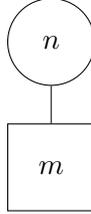
\begin{figure}[h!]
	\centering
	\begin{tikzpicture}[scale=1]
		\begin{scope}[auto, every node/.style={minimum size=3em,inner sep=1},node distance=0.5cm]
			\node[draw, circle] (v2) {$n$};
			\node[draw, below=of v2] (v1) {$m$};
		\end{scope}
		\draw(v1)--(v2);
	\end{tikzpicture}
	\caption{A linear quiver describing 3d $\CN=4$ SQCD with $n$ colors and $m$ flavors. The Higgs branch is identified (after resolving) with $T^*\textrm{Gr}(n,m)$. The action of $PSU(m)$ on $T^*\textrm{Gr}(n,m)$ satisfies the fundamental identity.}
	\label{fig:SQCD}
\end{figure}

In Section \ref{sec:CSYM} we introduce the Chern-Simons-matter theories of interest. We consider hypermultiplets valued in a symplectic vector space $R = \bigoplus_i T^*(V_i \otimes U_i)$ for $V_i \otimes U_i$ a product of representations for the gauge groups $G, H$, where the $\mathfrak{g}$ gauge fields belong to $\CN=4$ vector multiplets and the $\mathfrak{h}$-valued gauge fields belong to $\CN=2$ vector multiplets with Chern-Simons terms. The underlying gauge theory will not typically have $\CN=4$ supersymmetry, but is expected to flow to an $\CN=4$ theory in the IR so long as the fundamental identity is satisfied on the Higgs branch of the $\CN=4$ $G, R$ gauge theory. A fairly simple example of such a theory is $\CN=4$ $U(n)$ SQCD with $m \geq n$ flavors, \ie\, $G = U(n)$ and $V \otimes U = \C^n \otimes (\C^m)^* \cong \Hom(\C^n, \C^m)$, with quiver diagram in Fig. \ref{fig:SQCD}. The Higgs branch $T^* {\rm Gr}(n,m)$ has an $H = PSU(m)$ flavor symmetry that satisfies the fundamental identity, but it does not hold on $R = T^*(V \otimes U)$. See Section \ref{sec:SQCD} for more details.

A priori, such a theory will typically only have $\CN=4$ supersymmetry in the IR: in most cases of interest, the fundamental identity required for $\CN=4$ supersymmetry only holds on the Higgs branch of the gauged hypermultiplets. In particular, we should not expect a UV description of the theory to admit any topological twist. We are able to circumvent this issue by first passing to the $HT$-twisted theory described above, which exists for any 3d $\CN\geq 2$ theory. When the action of $H$ satisfies the fundamental identity on the Higgs branch, we find two topological deformations of the above $HT$-twisted theory and argue that the corresponding theories recover the $A$ (Section \ref{sec:CSYMA}) and $B$ (Section \ref{sec:CSYMB}) twists of the desired $\CN=4$ Gaiotto-Witten theory. In our analysis, we find that the $A$-twist deformation does not rely on the fundamental identity and find that the resulting theory is perturbatively topological, although we expect this to hold non-perturbatively for many theories of interest, in particular the $\CN=4$ Gaiotto-Witten theories. We explicitly describe the boundary VOAs in the $A$ and $B$ twist for some simple examples to illustrate their properties.

\subsubsection{$A$-twisted Gaiotto-Witten theories}
Relatively little is known about the $A$ twist of Gaiotto-Witten theories and there are few results to directly compare to our present work, although we hope our analysis will facilitate their study. The model we describe was first introduced in \cite{KLLtop}%
\footnote{The topological twist naming conventions used in \cite{KLLtop} are inverse to those used in the present paper. For example, our $A$-twisted theory corresponds to their $B$-model.}, %
where it was noted that hypermultiplet contributions to the action are $Q_A$-exact up to redefining the Chern-Simons gauge field; na{\"i}vely one might expect this to imply the hypermultiplets entirely decouple, but this is not the case.

\begin{figure}[h!]
	\centering
	\begin{tikzpicture}[scale=1]
		\begin{scope}[auto, every node/.style={minimum size=3em,inner sep=1},node distance=0.5cm]
			\node[draw, circle] (v2) {$N_k$};
			\node[draw, circle, right=of v2] (v1) {$M_{-k}$};
		\end{scope}
		\draw(v1)--(v2);
	\end{tikzpicture}
	\caption{A linear quiver describing the 3d $\CN=4$ Gaiotto-Witten theory corresponding to the supergroup $GL(N|M)$ at level $k$. The subscript ${}_k$ denotes a unitary gauge group realized by $\CN=2$ vector multiplets with Chern-Simons level $k$.}
	\label{fig:glNM}
\end{figure}
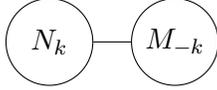

First consider hypermultiplets valued in a symplectic representation $R$ of a group $H$ gauged by an $\CN=2$ vector multiplet with non-vanishing Chern-Simons term at level $k$. For example, one could consider the theory in Fig. \ref{fig:glNM}, where $R = \Hom(\C^N, \C^M) \oplus \Hom(\C^M, \C^N)$ and $H_c = U(N) \times U(M)$, so $H = GL(N) \times GL(M)$. In Section \ref{sec:bdyCSYMA}, we find a boundary algebra given by the (derived) $H[\![z]\!]$ invariants of symplectic bosons valued in $R$ times auxiliary boundary degrees of freedom $\CV_{\rm aux}$ used to satisfy anomaly constraints:
\be
\label{eq:bdyCSAintro}
	\CV_A[R, H_k; \CV_{\rm aux}] = \big(\Sb[R] \times \CV_{\rm aux}\big)^{H[\![z]\!]}\,.
\ee
See \cite[Section 6.2]{CostelloDimofteGaiotto-boundary} for details about how to take derived invariants in the context of Neumann boundary conditions for gauge fields. Strictly speaking, this cannot be a VOA, \ie\, it cannot have a holomorphic stress tensor $T(z)$, if the derived $H[\![z]\!]$ invariants have support outside of degree zero.%
\footnote{The $c$ ghost used to take derived $H[\![z]\!]$ invariants has trivial OPEs with all other fields and this precludes the possibility of a holomorphic stress tensor.} %
This should not happen for the $\CN=4$ Gaiotto-Witten theories described above, as the existence of this holomorphic stress tensor is guaranteed by topological invariance of the bulk \cite{CostelloCreutzigGaiotto, CostelloDimofteGaiotto-boundary}. Nonetheless, as described below, we expect it should hold more generally. When the level of the bulk Chern-Simons field isn't too positive%
\footnote{One can work with levels that aren't too negative by a parity transformation, trading left boundary conditions for right boundary conditions and $k$ for $-k$. This is actually done in the recent paper \cite{CDGG} to more naturally match our VOA analysis.} %
we can choose the boundary degrees of freedom $\CV_{\rm aux}$ to be boundary Fermi multiplets transforming in a representation $N$ of $H$:
\be
\label{eq:bdyCSAintro2}
	\CV_A[R, H_k; N] = \big(\Sb[R] \times \Ff[N]\big)^{H[\![z]\!]}\,.
\ee
One particularly important aspect of the VOA $\Sb[R] \times \Ff[N]$ is that it has affine $\mathfrak{h}$ currents $J_{\mathfrak{h}}$ at level $-(k+h_H)$, where $h_H$ is the dual Coexeter number for $\mathfrak{h}$, that generate the $H[\![z]\!]$ action, \ie\, taking derived $H[\![z]\!]$ invariants can be identified with a derived version of a coset. When the derived invariants are concentrated in degree zero, we can then identify the boundary algebra with a coset VOA
\be
\label{eq:bdyCSAintrofin}
	\CV_A[R, H_k; N] = \frac{\Sb[R] \times \Ff[N]}{H_{-(k+h_H)}}\,.
\ee
Note that this boundary VOA doesn't use the fundamental identity in any way; we will return to this in more detail below.

When we include $\CN=4$ vector multiplets gauging a $G$ flavor symmetry of our hypermultiplets, we find a boundary VOA that combines the more general $A$-twist boundary VOAs $\CV_A[R, G]$ of Costello-Gaiotto with the above derived invariants. As a simple example, we can gauge the $H = PSU(m)$ flavor symmetry of $\CN=4$ SQCD presented by the quiver in Fig. \ref{fig:SQCD}. Concretely, the VOA $\CV_A[R, G]$ is realized as the $G$ BRST reduction of $R$-valued symplectic bosons together with an auxiliary VOA $\CV_{\rm aux}'$ to satisfy anomaly constraints. Just as above, when there is enough matter we can satisfy these anomaly constraints by coupling to boundary fermions in a representation $M$ of $G$:
\be
	\CV_A[R, G; M] = \big(\Sb[R] \times \Ff[M]\big) /\!/ G\,.
\ee
The appearance of a BRST reduction, as opposed to the above coset, is directly tied to using $\CN=4$ vector multiplets instead of $\CN=2$ vector multiplets with a Chern-Simons term. Combining the two types of gaugings is somewhat delicate; see Section \ref{sec:bdyCSYMA} for more details. A first approximation to this VOA is to take a coset of the product $\CV_A[R, G; M] \times \Ff[N]$
\be
\label{eq:bdyCSYMAapprox}
	\CV_A[R, G \times H_k; M, N] \approx \frac{\CV_A[R, G; M] \times \Ff[N]}{H_{-(k+h_H)}}\,.
\ee
In general, this is merely an approximation (roughly, the second page of a spectral sequence whose first page computes the BRST reduction $\CV_A[R, G; M] \times \Ff[N]$ and whose second page computes the derived $H[\![z]\!]$ invariants thereof) but seems to be rather close in many cases of interest. As such, the correct boundary VOA may involve a modification of the above coset, \eg\, by an orbifold, an extension, or combination thereof.

When there is good control of the Costello-Gaiotto boundary VOA $\CV_A[R, G; M]$, the above coset is a fairly manageable VOA. For example, the boundary VOA for $T[SU(n)]$, called the classical geometric Langlands kernel $A(\mathfrak{sl}(n))$, appears at corners in 4d $\CN=4$ super Yang-Mills \cite{CostelloGaiotto, CostelloCreutzigGaiotto, CreutzigGaiotto} and are central to the quantum Langlands program \cite{FrenkelGaiotto}. As advertised above, in the earlier work \cite{CDGG} we considered (a gauged linear $\sigma$-model incarnation of) the $\CN=4$ theory $T[SU(n)]$ and performed the above gauging for the $SU(n)$ Higgs-branch flavor symmetry. We argued that the corresponding $A$-twisted theory is a physical realization of the abstract 3d TQFTs of \cite{CGP, BCGPM, BGPMR}; properties of (a minor modification of) the above coset are central pieces of evidence supporting our expectations.

\subsubsection{``$A$ twist" of Chern-Simons-Yang-Mills theories}
One remarkable feature we uncover in our analysis of the $\CN=4$ Gaiotto-Witten theories is that the deformation from the $HT$ twist to the $A$ twist does not require the fundamental identity. When we consider theories that \emph{do not} satisfy the fundamental identity on the Higgs branch, we find a class of field theories obtained by gauging a Higgs-branch flavor symmetry $H$ of an $A$-twisted super Yang-Mills theory with $\CN=2$ Chern-Simons fields that are topological, at least perturbatively. The origin of this ``$A$-twist'' without the fundamental identity is somewhat mysterious, but seems to stem from the fact that the hypermultiplet scalars do not survive the $A$-twist and hence the fundamental identity, an equation imposed on these scalars, becomes trivialized.

The above analysis goes through without a hitch, but now for general $G, H, R$. For example, the boundary VOA for the ``$A$-twisted'' theory of hypermultiplets gauged by $\CN=2$ vector multiplets with non-vanishing Chern-Simons term is given by Eq. \eqref{eq:bdyCSAintro2}, without regard for the fundamental identity. Although we don't have a general proof of it, we expect that these derived invariants are concentrated in degree 0 and hence the boundary algebra can be identified with the coset VOA in Eq. \eqref{eq:bdyCSAintrofin}. The existence of the holomorphic stress tensor of this coset VOA is strong evidence that the bulk theory is quantum mechanically topological beyond perturbation theory.

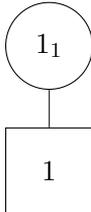
\begin{figure}[h!]
	\centering
	\begin{tikzpicture}[scale=1]
		\begin{scope}[auto, every node/.style={minimum size=3em,inner sep=1},node distance=0.5cm]
			\node[draw, circle] (v2) {$1_1$};
			\node[draw, below=of v2] (v1) {$1$};
		\end{scope}
		\draw(v1)--(v2);
	\end{tikzpicture}
	\caption{The quiver describing the 3d $\CN\neq4$ theory of $\CN=2$ $U(1)$ Chern-Simons gauge fields at level $1$ coupled to a single hypermultiplet.}
	\label{fig:U11plushyper}
\end{figure}

In Section \ref{sec:gaugedhyper} we consider $\CN=2$ $U(1)_1$ Chern-Simons theory coupled to a single hypermultiplet; see Fig. \ref{fig:U11plushyper}. The underlying supersymmetric gauge theory was argued in \cite{GLMMS} to flow to an $\CN=5$ theory in the IR, although the enhancement isn't obvious from our analysis as it is non-perturbative in nature. An argument identical to that appearing in \cite[Section 6.4.1]{CostelloDimofteGaiotto-boundary} shows that the boundary VOA is concentrated in cohomological degree zero, and we hence find that the corresponding ``$A$-twisted'' boundary VOA is the $U(1)_{-1}$-coset of a pair of symplectic bosons, realizing Kausch's $p=2$ singlet algebra $\mathfrak{M}(2)$ \cite{Kausch, WangWalgs}, also called Zamolodchikov's $\mathcal{W}_3$ algebra (at central charge $c=-2$) \cite{ZamoloWalgs}. This is one of the simplest logarithmic CFTs with much of its representation theory understood, see \eg\, \cite{CRlogCFT} and references therein. Its category of (finite-length) modules has a ribbon tensor structure \cite{CMYtensor1, CMYtensor2}, and the corresponding knot invariants built therefrom are related to the restricted unrolled quantum group at a fourth root of unity \cite{Murakami,GPMT,CMRunrolled} as well as indices of 3d $\CN=2$ theories $T[M_3]$ appearing in the 3d-3d correspondence \cite{DGG1,DGG2,3dmod,3foldVOA}. We compute half-indices \cite{GGP1, GGP2, DimofteGaiottoPaquette} counting local operators at the junction of bulk Wilson lines and this boundary condition, and identify them with characters of atypical singlet modules. It would be interesting to identify the bulk line operators corresponding to the remaining simple $\mathfrak{M}(2)$ modules. It is worth noting that the singlet VOA $\mathfrak{M}(2)$ arises as the $\psi \to \infty$ limit of the Gaiotto-Rap{\v c}{\'a}k corner VOA $Y_{1,1,0}[\psi]$ after decoupling a commutative subalgebra. The existence of such a deformation of the singlet $\mathfrak{M}(2)$ by a commutative subalgebra is directly analogous to that of the Feigin-Tipunin algebra described in \cite[Section 6.2]{CDGG} -- it should be interpreted as a background flat connection for a $U(1)$ flavor symmetry. This flavor symmetry deformation can also be interpreted as coming from a torus of the $SU(2)$ flavor symmetry of the symplectic fermion VOA described in \cite[Section 2]{CDGG}. It would be interesting to see if other examples of our ``$A$-twisted" theories lead to similar limits of other corner vertex algebras $Y_{L,M,N}[\psi]$.

\subsubsection{$B$-twisted Gaiotto-Witten theories}
Compared to the $A$ twist, much more is known about the $B$-twist of Gaiotto-Witten theories. For theories with linear targets, but without $\CN=4$ vector multiplets, \ie\, the Gaiotto-Witten theories associated to Lie supergroups $\widehat{H}$, the $B$-twisted theory can be identified with Chern-Simons theory based on the supergroup $\widehat{H}$ \cite{KapustinSaulina-CSRW} such as the one appearing in Fig. \ref{fig:glNM}. For more general hyperk\"ahler target $\CM$ (satisfying the fundamental identity), it can be interpreted as an AKSZ \cite{AKSZ} theory of maps into the graded symplectic manifold $\mathfrak{h}[1] \times \CM$ \cite{KQZ}. The $B$-twisted theory described in Section \ref{sec:CSYMB} is again an AKSZ theory with $\CM$ replaced by the Higgs branch $R/\!\!/G$, which is the natural gauge-theoretic replacement of $\CM$. We find that this AKSZ theory can also be viewed as a Chern-Simons theory based on the Lie superalgebra $(\mathfrak{h} \oplus \mathfrak{g} \oplus \mathfrak{g}^*) \oplus \Pi R$, where the first factor is the bosonic subalgebra and the second factor is the odd subspace, although it is fairly clear that this interpretation only makes sense perturbatively.

In Section \ref{sec:bdyCSYMB}, the boundary conditions we introduce admit boundary monopole operators, just as with $B$-twisted super Yang-Mills \cite{CostelloGaiotto}, and we only investigate the perturbative part of the boundary algebra. For the Gaiotto-Witten theory labeled by the supergroup $\widehat{H}$, we are able to identify the perturbative boundary algebra with a $\widehat{\mathfrak{h}}$ current algebra, as expected for Chern-Simons with gauge group $\widehat{H}$. It is natural to expect that the full, non-perturbative analysis results in the corresponding supergroup WZW model.

Including $\CN=4$ vector multiplets, such as in the theory obtained by gauging the $H = PSL(m)$ symmetry of the SQCD theory in Fig. \ref{fig:SQCD} that has $G = GL(N)$ and $R = \Hom(\C^N, \C^M) \oplus \Hom(\C^M, \C^N)$, yields a perturbative algebra of $(\mathfrak{h} \oplus \mathfrak{g} \oplus \mathfrak{g}^*) \oplus \Pi R$ currents. In this case, however, we should not expect the full, non-perturbative analysis to yield a supergroup WZW model: the physical theory only has monopole operators for the $G \times H$ factors, and hence one should only extend this perturbative current algebra by integrable modules (boundary monopole operators) for $G \times H$. This interpretation is supported by half-index computations, where one only sums over flux sectors for the physical gauge group. We hope to return to this problem in future work.

We end with the above mentioned SQCD example for $n = 1$, $m = 2$, \ie\, we gauge the $SU(2)$ Higgs-branch flavor symmetry of $\CN=4$ SQED with two hypermultiplets. As mentioned above, $\CN=4$ SQED with two hypermultiplets flows to $T[SU(2)]$ in the IR and we expect the bulk TQFT to exactly reproduce the $T^*\P^1/SU(2)_k$ Chern-Simons-Rozansky-Witten theory studied in \cite{KapustinSaulina-CSRW}. Once the non-perturbative aspects of this boundary algebra are better understood, we think that it would be fruitful to compare the category of modules of the extended VOA to the category of line operators proposed in \cite{KapustinSaulina-CSRW}.

\section{$\CN=4$ VOAs from Twisted Superfields}
\label{sec:twistedVOAs}
In this section we review results of \cite{twistedN=4} which are essential ingredients for our approach to describing algebras of boundary local operators in topologically twisted 3d theories. In particular, we review how to describe the $A$ and $B$ twists of traditional $\CN=4$ gauge theories using twisted superfields and how to derive the boundary VOAs of \cite{CostelloGaiotto} from this perspective.

Section \ref{sec:twistedN=2} is a lightning review of $HT$-twisted $\CN=2$ theories and the twisted superfields of \cite{AganagicCostelloMcNamaraVafa}, as presented by \cite{CostelloDimofteGaiotto-boundary}. In Section \ref{sec:N=4asN=2} we review the standard $\CN=4$ supermultiplets and describe the $HT$-twist of standard $\CN=4$ theories in the language of twisted superfields. The $A$- and $B$-twisted theories, as well as the $A$ and $B$ twist boundary VOAs of \cite{CostelloGaiotto}, are described in Section \ref{sec:SYMA} and Section \ref{sec:SYMB}, respectively.

\subsection{$HT$-twisted $\CN=2$ theories and twisted superfields}
\label{sec:twistedN=2}

Any three-dimensional $\CN=2$ supersymmetric field theory preserving the full $\CN=2$ $R$-symmetry $U(1)_R$ can be twisted to obtain a theory that behaves topologically in one spacetime direction and holomorphically in a plane transverse to that direction. The 3d $\CN=2$ supersymmetry algebra is generated by 2 spinors $Q_\alpha, \oQ_\alpha$, with the convention that $Q$ has $R$-charge $-1$ and $\oQ$ has $R$-charge 1, and non-trivial anti-commutation relations given by
\be
\{Q_\alpha, \oQ_\beta\} = (\sigma^\mu)_{\alpha \beta} P_{\mu},
\ee
in the absence of central charges. Up to symmetries of the algebra, there is a single nilpotent supercharge \cite{ESWtax, EagerSaberiWalcher}, which we take to be $Q_{HT} = \oQ_+$ and call the ``holomorphic-topological supercharge."

The $Q_{HT}$ twist is compatible with spacetimes that admit a ``transverse holomorphic foliation" (THF); they live on 3-manifolds that can locally be expressed as $\R_t \times \C_{z,\oz}$ with transition functions of the form
\be
	t \to t'(t,z,\oz) \qquad z \to z'(z) \qquad \oz \to \oz'(\oz)\,.
\ee
We only consider flat Euclidean space $\R^3 \cong \R_t \times \C_{z,\oz}$ or a Euclidean half-space $\R_{t \geq 0} \times \C_{z,\oz}$ in this paper. As long as we work on 3-manifolds with a THF, we lose nothing by considering a reduction of the 3d spin group ${\rm Spin}(3)$ to the 2d spin group ${\rm Spin}(2) \cong U(1)_{J_0}$ rotating vectors that are tangent to the leaves of the foliation. In flat space, taking the $Q_{HT}$ twist (also called the $HT$ twist) amounts to working with a modified spin group $U(1)_J$ generated by 
\be
	J = \tfrac{1}{2}R - J_0\,,
\ee
for which $Q_{HT}$ is a scalar, and then restricting attention to the $Q_{HT}$-cohomology of various observables.

The theories of interest are labeled by a (compact) gauge group $H_c$ with (real) Lie algebra $\mathfrak{h}_c$, a unitary representation $U$ thereof, an $H_c$-invariant holomorphic function $W: U \to \C$, and a collection of (possibly vanishing) levels $k$ whose quantization depends on the representation $U$. We denote by $U_r$ the subrepresentation of $R$-charge $r$. We will almost exclusively make use of the complexification of the gauge group $H_c$ (resp. Lie algebra $\mathfrak{h}_c$) and denote it $H$ (resp. $\mathfrak{h}$). To this data, we consider the $\CN=2$ Chern-Simons-matter theory of chiral multiplets valued in $U$ and vector multiplets valued in $\mathfrak{h}_c$ with superpotential $W$ and supersymmetric Chern-Simons terms at level $k$. We further require that we can assign integral or half-integral $U(1)_R$ $R$-charges to the chiral multiplets so that the superpotential $W$ has $R$-charge 2. See \cite{AHISS} for a more traditional analysis of the dynamics of these theories.

The twisted superfields of \cite{AganagicCostelloMcNamaraVafa} allow us to concisely formulate the $HT$ twist of the above $\CN=2$ Chern-Simons-matter theories. Schematically, one starts by reformulating the theory in the Batalin-Vilkovisky (BV) formalism \cite{BV}; the resulting theory has roughly twice as many fields, but this doubling allows the equations of motion and gauge symmetries to be encoded in a square-zero differential operator, the BV/BRST supercharge $Q_{\rm BV/BRST}$. To take the $HT$ twist of such a theory then corresponds to working in the cohomology of the combined differential $Q_{\rm BV/BRST} + Q_{HT}$ (as well as working with respect to the modified Lorentz group $U(1)_J$), which we also denote $Q_{HT}$ for notational simplicity. The twisted theory has two gradings. The first grading $C$ is cohomological and combines the BV/BRST ghost number ${\rm gh}$ and the $R$-charge $C = R + {\rm gh}$; the supercharge $Q_{HT}$ has $C = 1$. The second grading $J$ is via the modified Lorentz group $U(1)_J$, called ``twisted spin," and $Q_{HT}$ has charge $J = 0$.

In order to write down the $HT$-twisted theory of \cite{AganagicCostelloMcNamaraVafa}, we must set up some more notation. The THF of spacetime gives us a distinguished subspace of differential forms proportional to $\diff z$; we define $\bOmega^\bullet$ to be the quotient of all differential forms by those differential forms proportional to $\diff z$. The de Rham differential $\diff$ acting on differential forms induces a differential $\diff'$ acting on $\bOmega^\bullet$. Locally, we have $\bOmega^\bullet \cong C^\infty(\R^3)[\diff \oz, \diff t]$ and $\diff' = \pd_{\oz} \diff \oz + \pd_t \diff t$. Similarly, we can define $\bOmega^{\bullet, (j)}$ to be elements of $\bOmega^\bullet$ with coefficients the $j$-th power of the canonical bundle $K^j_{\C}$. Locally, we have $\bOmega^{\bullet, (j)} \cong C^\infty(\R^3)[\diff \oz, \diff t] \diff z^j$. There is a natural differential operator $\pd: \bOmega^{\bullet, (j)} \to \bOmega^{\bullet (j+1)}$ that locally takes the form $\pd = \pd_z \diff z$, a natural product $\bOmega^{\bullet, (j)} \otimes \bOmega^{\bullet, (j')} \to \bOmega^{\bullet, (j+j')}$ that treats $\diff z$ as Grassmann even and $\diff \oz, \diff t$ as Grassmann odd, and a natural integration map $\int: \bOmega^{2,(1)} \to \C$ of cohomological degree $-2$. The differential $\diff'$ acts as a odd derivation of this product, increases cohomology grading $C$ by 1, and leaves twisted spin $J$ unchanged. Similarly, $\pd$ acts as an even derivation of this product, increases twisted spin $J$ by 1, and leaves the cohomological grading $C$ unchanged.

The simplified model of \cite{AganagicCostelloMcNamaraVafa} contains the following physical, \ie\, ghost number 0, fields:
\begin{itemize}
	\item two components of the gauge field organized into the fermionic field $$A = A_t \diff t + A_\oz \diff \oz \in \bOmega^{1, (0)} \otimes \mathfrak{h}[1],$$ with $A_t$ complexified by the real scalar $\sigma$ of the $\CN=2$ vector multiplet;
	\item a bosonic field $$B = B_z \diff z \in \bOmega^{0, (1)} \otimes \mathfrak{h}^*$$ identified in the physical theory with the curvature $\tfrac{1}{g^2} F_{zt}$ up to terms depending on the Chern-Simons levels;
	\item a $U$-valued bosonic field $$\phi = \sum\limits_r \phi_{r}, \quad  \phi_{r} = \phi_{r,z} \diff z^{r/2} \in \bOmega^{0,(r/2)} \otimes U_r$$ identified with the bosons in the chiral superfields after applying the twisting homomorphism turning scalars of $R$-charge $R = r$ into sections of $K_{\C}^{r/2}$;
	\item a $U^*$-valued fermionic field $$\eta = \sum\limits_r \eta_{r}, \quad \eta_{r} = \big(\eta_{r,t} \diff t + \eta_{r,\oz} \diff \oz \big) \diff z^{1-r/2} \in \bOmega^{1,(1-r/2)} \otimes U_r^*[1],$$ whose components are identified with the covariant derivatives of the conjugate scalar $\eta_t \sim \overline{D_\oz \phi}, \eta_\oz \sim \overline{D_t \phi}$.
\end{itemize}
In these expressions, $[1]$ denotes a shift in cohomological degree; for example, $A$ is a 1-form of cohomological degree $C=1$ (the components $A_t, A_z$ have vanishing ghost number and $R$-charge; the differentials $\diff t, \diff \oz$ are given ghost number 1 and vanishing $R$-charge) and twisted spin $J=0$, whereas $B$ is a 1-form of cohomological degree $C=0$ and twisted spin $J=1$. In addition to the above fields, there is the usual BRST ghost $c \in \bOmega^{0, (0)} \otimes \mathfrak{h}[1]$ for the $H_c$ gauge symmetry; an ``exotic" ghost $\psi = \sum_r \psi_r$ for the action of the $HT$-supercharge $Q_{HT}$ with $\psi_r  \in \bOmega^{0, (1-r/2)} \otimes U_r^*[1]$ corresponding to a fermion in the anti-chiral multiplet conjugate to $\phi_r$; and various anti-fields $A^*, B^*, \phi^*, \eta^*, c^*, \psi^*$. The beauty of this model is that these fields, ghosts, and anti-fields can be collected into ``twisted superfields" having homogeneous cohomological degree but inhomogeneous form degree:
\be
\begin{aligned}
	\label{eq:superfields}
	\bA & = c + A + B^* \in \bOmega^{\bullet, (0)} \otimes \mathfrak{h}[1] \qquad & \bB & = B + A^* + c^* \in \bOmega^{\bullet, (1)} \otimes \mathfrak{h}^*\\
	\bPhi_r & = \phi_r + \eta_r^* + \psi_r^* \in \bOmega^{\bullet, (r/2)} \otimes U_r \qquad & \bPsi_r & = \psi_r + \eta_r + \phi_r^* \in \bOmega^{\bullet, (1-r/2)} \otimes U_r^*[1]
\end{aligned}\,.
\ee
The cohomological and twisted spin gradings of these fields are listed in Table \ref{table:HTspincoho}.

\begin{table}[h!]
	\centering
	\begin{tabular}{c|c|c|c|c}
		& $\bA$   & $\bB$  & $\bPhi_r$ & $\bPsi_r$ \\ \hline
		$(J, C)$ & $(0,1)$ & $(1,0)$ & $(r/2,r)$ & $(1-r/2,1-r)$
	\end{tabular}
	\caption{Twisted spin ($J$) and cohomological grading ($C$) of various twisted superfields. The twisted superfields $(\bA, \bB)$ originate from $\CN=2$ vector multiplets and $(\bPhi_r, \bPsi_r)$ originate from $\CN=2$ chiral multiplets of $R$-charge $r$.}
	\label{table:HTspincoho}
\end{table}

In terms of twisted superfields, the classical action is given by
\be
	S = \int \bB F'(\bA) + \tfrac{k}{4\pi} \bA \pd \bA + \bPsi \diff'_{\bA} \bPhi + \bW\,,
\ee
where $\diff'_{\bA} = \diff' + \bA$ is the exterior covariant derivative with curvature $F'(\bA) = \diff' \bA + \bA^2$, and $\bW = W(\bPhi)$ is the (pull-back of the) superpotential. The action of $Q_{HT}$ is given by
\be
\begin{aligned}
	\label{eq:Q}
	Q_{HT} \bA & = F'(\bA) \qquad & Q_{HT} \bB & = \diff'_{\bA} \bB - \bmu + \tfrac{k}{2\pi} \pd \bA\\
	Q_{HT} \bPhi & = \diff'_{\bA} \bPhi & Q_{HT} \bPsi & = \diff'_{\bA} \bPsi + \frac{\pd \bW}{\pd \bPhi}\\
\end{aligned}\,,
\ee
where $\bmu = \mu(\bPhi, \bPsi)$ is the (pull-back of the) moment map for the action of $H$ on $T^*[1]U$. The supercharge $Q_{HT}$ can be viewed as the Hamiltonian vector field associated to the Hamiltonian $S$, \ie\,  $Q_{HT} = \{-,S\}_{\rm BV}$, with respect to the BV brackets given by
\be
	\{\bA(x), \bB(y)\}_{\rm BV} = \delta(x-y) \diff{\rm Vol} = \{\bPhi(x), \bPsi(y)\}_{\rm BV}\,.
\ee
The results of \cite{GwilliamWilliams} show that any of the above classical theories admits a finite, 1-loop exact, perturbative quantization.

\subsection{$\CN=4$ theories as $\CN=2$ theories}
\label{sec:N=4asN=2}
We now apply the above twisted superfields to $\CN=4$ theories of hypermultiplets gauged by $\CN=4$ vector multiplets. The 3d $\CN=4$ supersymmetry algebra has an $\Spin(4)_R \simeq SU(2)_A \times SU(2)_B$ $R$-symmetry and is generated by 4 spinors $Q^{a \dot{a}}_\alpha$, where $a$ is an $SU(2)_A$ doublet index and $\dot{a}$ is an $SU(2)_B$ doublet index so that the supercharges transform in the vector representation of $\Spin(4)_R$, with non-trivial anti-commutation relations given by
\be
	\{Q^{a \dot{a}}_\alpha, Q^{b \dot{b}}_\beta\} = \epsilon^{ab} \epsilon^{\dot{a} \dot{b}} (\sigma^\mu)_{\alpha \beta} P_{\mu},
\ee
in the absence of central charges. Up to symmetries of the algebra, there is a single holomorphic-topological supercharge $Q_{HT}$ and two topological supercharges $Q_A, Q_B$ \cite{ESWtax, EagerSaberiWalcher} that we take to be
\be
	Q_{HT} = Q^{+ \dot{+}}_+ \qquad Q_A = \delta^\alpha{}_{a} Q^{a \dot{+}}_\alpha = Q^{+ \dot{+}}_+ + Q^{- \dot{+}}_-  \qquad Q_B = \delta^\alpha{}_{\dot{a}} Q^{+ \dot{a}}_\alpha = Q^{+ \dot{+}}_+ + Q^{+ \dot{-}}_-.
\ee

If we write $Q^1 = Q^{-\dot{-}},\, \oQ^1 = Q^{+\dot{+}},\, Q^2 = - Q^{-\dot{+}},$ and $\oQ^2 = Q^{+\dot{-}}$, the above supersymmetry algebra takes the form
\be
	\{Q^i_\alpha, \oQ^j_\beta\} = \delta^{ij} (\sigma^\mu)_{\alpha \beta} P_{\mu}.
\ee
This splitting leaves manifest a $U(2) \hookrightarrow SU(2)_A \times SU(2)_B$ $R$-symmetry. This $U(2)$ has a maximal torus given by the product of maximal tori $U(1)_A \times U(1)_B \hookrightarrow SU(2)_A \times SU(2)_B$. The choice of maximal torus $U(1)_A$ (resp. $U(1)_B$) is equivalent to a choice of complex structure on the Higgs branch (resp. Coulomb branch).

We will work with respect to the $\CN=2$ subalgebra generated by $Q^1, \oQ^1$, which contains $Q_{HT} = \oQ^1_+$. One natural choice of $R$-symmetry group is the diagonal $U(1)_R \hookrightarrow U(1)_A \times U(1)_B$; the supercharges $Q^1, \oQ^1$ then have charges $-1, 1$ while $Q^2, \oQ^2$ are both chargeless.%
\footnote{Although it is highly symmetric, \eg\, it is preserved by the 3d mirror automorphism, this choice if $R$-charge is not well-suited for placing the $HT$-twisted theory on curved spacetimes. The issue is that the properly normalized $R$-symmetry generator, \ie\, the minimal choice so that a rotation by $2\pi$ acts trivially on all fields, is twice this generator. Correspondingly, hypermultiplets have half-integral $U(1)_R$ charge and, as we will see, twisting will require a choice of fourth root of the canonical bundle of the ``spatial'' surface $\Sigma$, \ie\, a choice of square root $K^{1/4}_\Sigma = (K^{1/2}_\Sigma)^{1/2}$ of the spin structure $K^{1/2}_\Sigma$. This isn't an issue when formulating the $HT$-twisted theory on flat space, but such fourth roots do not exist in general, \eg\, for the sphere $\Sigma = S^2$. In this paper we mostly ignore this point as we focus on flat space and the topologically theories which use a different choice of $R$-symmetry that doesn't suffer from this issue. See \cite[Section 2.2]{twistedN=4} for more details.} %
The anti-diagonal $U(1)_F \hookrightarrow U(1)_A \times U(1)_B$ acts trivially on $Q^1, \oQ^1$ and will be realized as an additional flavor symmetry of the twisted theory.

Denote the (complexified) gauge group $G$%
\footnote{We use the letter $G$ to denote the complexified gauge group associated to the Yang-Mills gauge fields residing in $\CN=4$ vector multiplets, and use $H$ for the complexified gauge group associated to the Chern-Simons gauge fields. We also reserve the letter $V$ for a complex representation of $G$ and reserve $U$ for a complex representation of $H$.} %
and assume that the hypermultiplets transform in a complex symplectic representation of the form $R = T^*V := V \oplus V^*$ for $V$ some complex representation of $G$. With respect to our chosen $\CN=2$ subalgebra, the basic $\CN=4$ multiplets, namely, hypermultiplets and vector multiplets, decompose as follows. The hypermultiplet decomposes into two chiral/anti-chiral pairs with one chiral valued in $V$ and the other valued in $V^*$. We will denote the corresponding twisted superfields $(\bX, \bPsi_{\bX})$ and $(\bY, \bPsi_{\bY})$; based on the above choice of $\CN=2$ subalgebra, we find that the $R$-charges of $X$ and $Y$ are given by 
\be
	r_X = r_Y = \tfrac{1}{2}\,.
\ee
Similarly, an $\CN = 4$ vector multiplet decomposes into an $\CN = 2$ vector multiplet and a chiral/anti-chiral pair valued in the adjoint representation $\mathfrak{g}$. We will denote the corresponding twisted superfields by $(\bA, \bB)$, for the $\CN=2$ vector multiplet, and $(\bPhi, \bLambda)$, for the $\CN=2$ chiral/anti-chiral pair. The $\CN=2$ $R$-charge of the chiral multiplet scalar $\phi$ is given by
\be
	r_{\phi} = 1\,.
\ee

Written as an $\CN=2$ theory, the above $\CN=4$ theory has a superpotential of the form $W = -Y \phi X$. We can therefore write the $HT$-twisted action of this $\CN=4$ theory as
\be
\label{eq:twistedactionSYM}
	S = \int \bB F'(\bA) + \bLambda \diff'_{\bA} \bPhi+ \bPsi_{\bX} \diff'_{\bA} \bX + \bPsi_{\bY} \diff'_{\bA} \bY - \bY \bPhi \bX,
\ee
and thus the action of $Q_{HT}$ is given by 
\be
\begin{aligned}
	\label{eq:QSYM}
	Q_{HT} \bA & = F'(\bA) \qquad & Q_{HT} \bB & = \diff'_{\bA} \bB - \bmu\\
	Q_{HT} \bPhi & = \diff'_{\bA} \bPhi & Q_{HT} \bLambda & = \diff'_{\bA} \bLambda - \bmu_\C\\
	Q_{HT} \bX & = \diff'_{\bA} \bX \qquad & Q_{HT} \bPsi_{\bX} & = \diff'_{\bA} \bPsi_{\bX} - \bY \bPhi\\
	Q_{HT} \bY & = \diff'_{\bA} \bY \qquad & Q_{HT} \bPsi_{\bY} & = \diff'_{\bA} \bPsi_{\bY} - \bPhi \bX\\
\end{aligned},
\ee
where $\bmu_\C = \mu_\C(\bX, \bY)$ is the moment map for the $\mathfrak{g}$ action on $T^*V$.

There are two particularly important flavor symmetries present for any choice of $G$ and $V$. First, there is a $U(1)_M$ flavor symmetry%
\footnote{This rotation of the hypermultiplets may be actually be gauged, \eg\, in a $G_c = U(n)$ gauge theory with $m$ fundamental hypermultiplets $V = (\C^n)^{\oplus m} \cong \Hom(\C^m, \C^n)$. Nonetheless, it is guaranteed to commute with any the gauge symmetry and hence can be use to redefine the cohomological degree.} %
rotating the $V$ and $V^*$ chiral multiplets of the hypermultiplet with opposite phases and leaving the $\CN=4$ vector multiplet unchanged. We will need to use this symmetry to ensure that the bosons in these multiplets have even cohomological grading in the $B$-twist, \cf\, Section 5.2.3 of \cite{descent}; the corresponding charges are given in Table \ref{table:flavM}.%
\footnote{One should perform a similar regrading using the topological flavor symmetry measuring monopole charge to ensure that monopole operators have even cohomological degree in the $A$ twist, although we will mostly ignore this point.} 
Additionally, there is a $U(1)_F$ symmetry coming from the $\CN = 4$ $R$-symmetry group. The $U(1)_F$ charges of the twisted superfields coming from the $\CN=4$ multiplets are given in Table \ref{table:flavF}.

\begin{table}[h!]
	\centering
	\begin{tabular}{c|c|c|c|c}
		& $\bX$ & $\bPsi_{\bX}$ & $\bY$ & $\bPsi_{\bY}$\\ \hline
		$M$ & $1$   & $-1$    & $-1$ & $1$\\
	\end{tabular}
	\caption{Charges under the $U(1)_M$ flavor symmetry rotating the $\CN=4$ hypermultiplets written in the language of twisted superfields.}
	\label{table:flavM}
\end{table}
\begin{table}[h!]
	\centering
	\begin{tabular}{c|c|c|c|c|c|c|c|c}
		& $\bX$ & $\bPsi_{\bX}$ & $\bY$ & $\bPsi_{\bY}$ & $\bA$   & $\bB$  & $\bPhi$ & $\bLambda$\\ \hline
		$F$ & $1$   & $-1$    & $1$ & $-1$ & $0$ & $0$ & $-2$ & $2$\\ 
	\end{tabular}
	\caption{Charges under the residual $R$-symmetry, viewed as a flavor symmetry, of $\CN=4$ multiplets written in the language of twisted superfields. This corresponds to the anti-diagonal $U(1)$ in $U(1)_A \times U(1)_B$.}
	\label{table:flavF}
\end{table}

\subsection{Topologically twisted $\CN=4$ theories}
We now wish to deform the above $HT$-twisted theory to its topological $A$ and $B$ twists. As the action $S$ and the differential $Q$ are tied to one another in the BV formalism, these deformations are realized by adding new terms to the twisted actions above. Using knowledge of the action of the full $\CN=4$ superalgebra and how to deform $Q_{HT}$ to $Q_A$ and $Q_B$, this is a relatively straightforward task. 

It is worth noting that it is possible to combine the twisted superfields coming from these $\CN=4$ multiplets as
\be
\label{eq:combined}
\begin{aligned}
	\CX & = \bX - \varepsilon \bPsi_{\bY} \qquad & \CY & = \bY + \varepsilon \bPsi_{\bX}\\
	\CA & = \bA + \varepsilon \bPhi \qquad & \CB & = \bLambda + \varepsilon \bB\\
\end{aligned}\,,
\ee
where $\varepsilon$ is an odd parameter with $U(1)_J \times U(1)_C \times U(1)_F \times U(1)_M$ charge $(-\tfrac{1}{2}, 0, 2, 0)$, from which it follows that the $HT$ twist of super Yang-Mills is an AKSZ theory \cite{AKSZ} based on the mapping space
\be
\label{eq:EOMYMHTSYM}
\Sect(\R_\dR \times \C_\opd \times \C^{0|1}, T^*(V/G) \otimes K^{1/4}_\C)\,.
\ee
The deformation of the $HT$ twist to the topological $A$ twist (resp. $B$ twist) is reproduced by deforming the differential on this mapping space by $\pd_\varepsilon$ (resp. $- \varepsilon \pd$) \cite{CostelloPSI, Butson2, ESWtax}. In terms of the space of solutions to the equations of motion, the deformations take the following form.
\begin{figure}[H]
	\centering
	\begin{tikzpicture}
		\draw (0,1.5) node {$\Sect(\R_\dR \times \C_\opd \times \C^{0|1}, T^*(V/G) \otimes K^{1/4}_\C)$};
		\draw (-3,-0.25) node {$\Sect(\R_\dR \times \C_\opd, (T^*(V/G) \otimes K^{1/2}_\C)_{\dR}$};
		\draw (2.5,-0.25) node {$\Maps(\R^3_\dR, T^*[2](V/G))$};
		\draw[->,decorate,decoration={snake,amplitude=.4mm,segment length=3mm, post length=1mm}] (-1.25,1.25) -- (-1.75,0.25);
		\draw (-2.5, 0.75) node {$A$ twist};
		\draw[->,decorate,decoration={snake,amplitude=.4mm,segment length=3mm, post length=1mm}] (1.25,1.25) -- (1.75,0.25);
		\draw (2.5, 0.75) node {$B$ twist};
	\end{tikzpicture}
\end{figure}

\subsubsection{$A$ twist}
\label{sec:SYMA}
First consider deforming the $HT$ twist to the topological $A$ twist. The topological supercharge $Q_A$ is given by the sum of $Q_{HT} = \oQ^1_+$ and $-Q^2_-$. We also will need to mix the twisted spin and cohomological grading with $U(1)_F$ charge to ensure that $Q_A$ is a scalar and of cohomological degree 1. Since $Q_{HT}$ has charges $(0,1,0)$ under $U(1)_J \times U(1)_C \times U(1)_F$ and $Q^2_-$ has charges $(\tfrac{1}{2}, 0, -2)$, the appropriate mixing corresponds to
\be
\label{eq:Amixing}
	J_A = J + \tfrac{1}{4} F \hspace{1cm} C_A = C - \tfrac{1}{2}F.
\ee
The new twisted spin and cohomological grading of our fields are given in Table \ref{table:Aspincoho}.

\begin{table}[h!]
	\centering
	\begin{tabular}{c|c|c|c|c|c|c|c|c}
		& $\bA$   & $\bB$  & $\bPhi$ & $\bLambda$ & $\bX$ & $\bPsi_{\bX}$ & $\bY$ & $\bPsi_{\bY}$ \\ \hline
		$(J_A, C_A)$ & $(0,1)$ & $(1,0)$ & $(0,2)$ & $(1,-1)$ & $(\tfrac{1}{2},0)$   & $(\tfrac{1}{2},1)$    & $(\tfrac{1}{2},0)$ & $(\tfrac{1}{2},1)$ 
	\end{tabular}
	\caption{Twisted spin ($J_A$) and cohomological grading ($C_A$) of the twisted superfields in the topological $A$ twist of super Yang-Mills.}
	\label{table:Aspincoho}
\end{table}

We can identify the deformation of the above $HT$-twisted theory to an $A$-twisted theory from the supersymmetry transformations in the physical theory.  Upon investigating the transformation of the hypermultiplets under $Q^2_-$, one finds that $Q^2_- \psi_X = Q^2_- \psi_Y = 0,$ $Q^2_- X \sim \psi_Y,$ and $Q^2_- Y \sim \psi_X.$ Similarly, the vector multiplet fields transform as, $Q^2_- (F_{zt} + i D_z \sigma) = Q^2_- \varphi = 0,$ $Q^2_- (A_t - i \sigma) \sim \lambda_+$ and $Q^2_- \lambda_- \sim F_{zt} + i D_z \sigma$. We see that the above $HT$-twisted theory is deformed to the $A$-twist by deforming the action to
\be
\label{eq:AtwistedactionSYM}
S_A = S + \int\bB \bPhi - \bPsi_{\bX} \bPsi_{\bY}\,.
\ee
The resulting action of $Q_A$ is given by
\be
\begin{aligned}
	\label{eq:QASYM}
	Q_A \bA & = F'(\bA) + \bPhi \qquad & Q_A \bB & = \diff'_{\bA} \bB - \bmu\\
	Q_A \bPhi & = \diff'_{\bA} \bPhi & Q_A \bLambda & = \diff'_{\bA} \bLambda - \bmu_\C + \bB\\
	Q_A \bX & = \diff'_{\bA} \bX - \bPsi_{\bY} \qquad & Q_A \bPsi_{\bX} & = \diff'_{\bA} \bPsi_{\bX} + \bY \bPhi\\
	Q_A \bY & = \diff'_{\bA} \bY + \bPsi_{\bX} \qquad & Q_A \bPsi_{\bY} & = \diff'_{\bA} \bPsi_{\bY} + \bPhi \bX\\
\end{aligned}.
\ee
As expected, the difference $Q_A - Q_{HT}$ exactly reproduces $\pd_\epsilon$ acting on the recombined fields in Eq. \eqref{eq:combined}.

Given the above twisted description for our $A$-twisted gauge theory, it is relatively straightforward to recover the boundary VOA of \cite{CostelloGaiotto}. We consider the same basic boundary conditions: $\CN=(0,4)$ Neumann boundary conditions for the $\CN=4$ vector multiplets, \ie\, $(0,2)$ Neumann for the $\CN=2$ vector multiplet and $(0,2)$ Dirichlet for the adjoint chiral multiplet, as well as the $\CN=4$ hypermultiplets, \ie\, $(0,2)$ Neumann boundary conditions for both $V$ and $V^*$ chiral multiplets. Additionally, one must include boundary degrees of freedom to cancel any gauge anomalies for the boundary gauge symmetry.

With these $\CN=(0,4)$ Neumann boundary conditions one finds, in the conventions of \cite{DimofteGaiottoPaquette, CostelloDimofteGaiotto-boundary}, that a vector multiplet contributes $-2h_G$ to the anomaly coefficient and a hypermultiplet contributes $\tfrac{1}{2}T_{T^*V} = T_V$, where $h_G$ is the dual Coexeter number of $\mathfrak{g}$ and $T_V$ is the quadratic index of $V$ normalized to $2h_G$ in the adjoint representation, whence we must include boundary degrees of freedom contributing $T_V - 2h_G$ to the anomaly coefficient. For many cases of interest, we have $T_V \geq 2h_G$ and can introduce a boundary Fermi multiplet transforming in a representation $M$ satisfying $T_M = T_V - 2 h_G$. In terms of the above twisted formalism, we introduce boundary twisted superfields $\bGamma, \wt{\bGamma}$ for the $M$-valued Fermi multiplet, with boundary action
\be
	S_\pd = \int_\pd \wt{\bGamma} \opd_{\bA} \bGamma\,.
\ee
The bulk twisted superfields are required to satisfy the following boundary conditions:
\begin{itemize}
	\item Neumann boundary conditions for the $\CN = 2$ vector multiplet. ($\bB|_{\pd} =  \bmu_\pd$)
	\item Dirichlet boundary conditions for the adjoint chiral multiplet. ($\bPhi|_{\pd} = 0$)
	\item Neumann boundary conditions for the hypermultiplets. ($\bPsi_{\bX}|_{\pd}, \bPsi_{\bY}|_{\pd} = 0$)
\end{itemize}
The boundary value of $\bB$ is simply the moment map/current $\bmu_\pd$ for the $\mathfrak{g}$ action on the boundary fields. 

Importantly, the Neumann boundary conditions on the vector multiplets ensure that there are no boundary monopole operators and thus it suffices to consider a perturbative analysis. As described in \cite{twistedN=4}, before taking into account gauge invariance, one finds an algebra of boundary local operators generated by the lowest components $c, \lambda, X, Y$ of the bulk twisted superfields $\bA, \bLambda, \bX, \bY$ and the lowest components $\gamma, \wt{\gamma}$ of the boundary twisted superfields $\bGamma, \wt{\bGamma}$. One finds that the $\oz$ dependence of these operators is trivial in cohomology and that the only non-trivial OPEs of these fields are given by
\be
\label{eq:SYMAope}
	c^a(z) \lambda_b(w) \sim \frac{\delta^a{}_b}{z-w} \qquad X^n(z) Y_m(w) \sim \frac{\delta^n{}_m}{z-w} \qquad \gamma^p(z) \wt{\gamma}_q(w) \sim \frac{\delta^p{}_q}{z-w}\,,
\ee
where $a,b$ are Lie algebra indices, $n,m$ are indices for the representation $V$ and its dual $V^*$, and $p,q$ are indices for the representation $M$ and its dual $M^*$. The residual action of $Q_A$ on these operators is non-zero and encodes the action of holomorphic gauge transformations
\be
\label{eq:YMAQ}
\begin{aligned}
	Q_A c & = c^2 \qquad & Q_A \lambda & = J\\
	Q_A X & = c \cdot X \qquad & Q_A Y & = c \cdot Y\\
	Q_A \gamma & = c \cdot \gamma \qquad & Q_A \wt{\gamma} & = c \cdot \wt{\gamma}\\
\end{aligned}\,,
\ee
where $J$ is the current generating the affine $\mathfrak{g}$ action; in components it reads
\be
	J_a = f^c{}_{ab} \norm{\lambda_c c^b} + (\tau_a)^m{}_n \norm{Y_m X^n} + (\sigma_a)^p{}_q \norm{\wt{\gamma}_p \gamma^q}\,.
\ee
As described in detail in \cite[Section 6.2.1]{CostelloDimofteGaiotto-boundary}, the precise way one takes gauge invariants must be done by first removing the zero mode of $c$, then taking $Q_A$ cohomology of the (constant) $G$ invariants in the algebra generated by $\pd_z c(z), \lambda(z), X(z), Y(z), \gamma(z), \wt{\gamma}(z)$ and their $\pd_z$ derivatives.%
\footnote{This somewhat ad hoc looking procedure is natural from the perspective of derived invariant theory. In such a setting, one finds that taking invariants with respect to a reductive group $G$, \eg\, the complexification of a compact, semisimple Lie group, is an exact functor. This should be contrasted with taking invariants with respect to a Lie algebra $\mathfrak{g}$, which leads to the notion of Lie algebra (co)homology. When taking derived invariants for a group like $G[\![z]\!]$, with $G$ reductive, the non-constant transformations behave like the Lie algebra $z \mathfrak{g}[\![z]\!]$ and one should include the corresponding modes of the $c$ ghost. Invariants for the constant transformations, on the other hand, should be taken without the use of a ghost.} %
We therefore find that the boundary algebra is simply the $G$-BRST reduction of $T^*V$-valued symplectic bosons (generated by $X,Y$) times $M$-valued complex fermions (generated by $\gamma, \wt{\gamma}$), exactly reproducing the boundary algebra of \cite{CostelloGaiotto}.

\subsubsection{$B$ twist}
\label{sec:SYMB}
We now turn to the $B$ twist. The topological supercharge $Q_B$ is given by the sum of $Q_{HT}$ with $\oQ^2_-$. Again, we will need to mix the twisted spin and cohomological grading with $U(1)_F$ charge to ensure that $Q_B$ is a scalar and of cohomological degree 1. Since $\oQ^2_-$ has charges $(\tfrac{1}{2},0,1)$ under $U(1)_J \times U(1)_C \times  U(1)_F$, the appropriate mixing corresponds to
\be
\label{eq:Bmixing}
	J_B = J - \tfrac{1}{2} F \hspace{1cm} C_B = C + F - M
\ee
The new twisted spin and cohomological gradings are given in Table \ref{table:Bspincoho}.

\begin{table}[H]
	\centering
	\begin{tabular}{c|c|c|c|c|c|c|c|c}
		& $\bA$   & $\bB$  & $\bPhi$ & $\bLambda$ & $\bX$ & $\bPsi_{\bX}$ & $\bY$ & $\bPsi_{\bY}$ \\ \hline
		$(J_B, C_B)$ & $(0,1)$ & $(1,0)$ & $(1,0)$ & $(0,1)$ & $(0,0)$   & $(1,1)$    & $(0,2)$ & $(1,-1)$ 
	\end{tabular}
	\caption{Twisted spin ($J_B$) and cohomological grading ($C_B$) of the twisted superfields in the topological $B$ twist of super Yang-Mills.}
	\label{table:Bspincoho}
\end{table}

Again, we expect that the deformations for the hypermultiplets and vector multiplets are decoupled from one another. Upon investigating the transformation of the hypermultiplets under $\oQ^2_-$, one finds that $\oQ^2_- X = \oQ^2_- Y = 0,$ $\oQ^2_- \psi_X \sim \pd_z Y,$ and $\oQ^2_- \psi_Y \sim \pd_z X.$ Similarly, one finds $\oQ^2_- (A_t -i \sigma) = \oQ^2_- \lambda_+ = 0$, $\oQ^2_- (F_{zt} + i D_z \sigma) \sim \pd \lambda_+$ and $\oQ^2_- \varphi \sim \pd \lambda_-$. Using the fact that the ($z$ covariant derivative of the) $c$ ghost is cohomologous to $\lambda_-$ and $F_{zt} + i D_z \sigma$ is cohomologous to $B$, the deformation of the action takes the form
\be
\label{eq:BtwistedactionYM}
	S_B = S + \int\bY \pd \bX - \bLambda \pd \bA
\ee
which indeed has the desired charges. The action of the supercharge $Q_B$ is then given by
\be
\begin{aligned}
	\label{eq:QBYM}
	Q_B \bA & = F'(\bA) \qquad & Q_B \bB & = \diff'_{\bA} \bB - \bmu - \pd \bLambda\\
	Q_B \bPhi & = \diff'_{\bA} \bPhi - \pd \bA & Q_B \bLambda & = \diff'_{\bA} \bLambda - \bmu_\C\\
	Q_B \bX & = \diff'_{\bA} \bX \qquad & Q_B \bPsi_{\bX} & = \diff'_{\bA} \bPsi_{\bX} + \bY \bPhi - \pd \bY\\
	Q_B \bY & = \diff'_{\bA} \bY \qquad & Q_B \bPsi_{\bY} & = \diff'_{\bA} \bPsi_{\bY} + \bPhi \bX + \pd \bX\\
\end{aligned}.
\ee
In particular, the difference $Q_B - Q_{HT}$ is the expected $-\epsilon \pd$.

Before moving on to the boundary VOA of \cite{CostelloGaiotto}, we note that if we combine the fields as in Eq. \eqref{eq:combined} the classical action takes a remarkably simple form:
\be
\label{eq:BtwistedactionYM2}
	S_B = \int \CB F(\CA) + \CY \diff_\CA \CX\,.
\ee
This action exactly matches the action of a Chern-Simons theory based off of the Lie superalgebra $T^*\mathfrak{g} \oplus \Pi(T^*V)$ with non-trivial brackets
\be
\label{eq:SYMBliealg}
\begin{alignedat}{3}\relax
	[T_a, T_b] & = f^c{}_{ab} T_c &&&  [T_a, S^b] & = f^b{}_{ac} S^c\\
	[T_a, \theta_m] & = \theta_n (\tau_a)^n{}_m &&&  [T_a, \otheta^n] & = -(\tau_a)^n{}_m \otheta^m\\
	&& \hspace{-1.5cm} \{\theta_m, \otheta^n\} & = (\tau_a)^n{}_m S^a \hspace{-1.5cm} &&\\
\end{alignedat}\,,
\ee
where the $T_a$ are a basis of $\mathfrak{g}$, the $\theta_n$ are a basis of $\Pi V$, and the $S^a$ and $\otheta^n$ are dual bases of $\mathfrak{g}^*$ and $\Pi V^*$. Equivalently, this theory corresponds to the an AKSZ theory based on maps into the Higgs branch $T^*[2] (V/G)$.

Just as with the $A$ twist, the above $B$-twisted theory allows us to reproduce the boundary VOA of \cite{CostelloGaiotto}. We consider the following boundary conditions:
\begin{itemize}
	\item Dirichlet boundary conditions for the $\CN = 2$ vector multiplet. ($\bA|_{\pd} = 0$)
	\item Neumann boundary conditions for the adjoint chiral multiplet. ($\bLambda|_{\pd} = 0$)
	\item Dirichlet boundary conditions for the hypermultiplets. ($\bX|_{\pd}, \bY|_{\pd} = 0$)
\end{itemize}
This $\CN=(0,4)$ Dirichlet boundary condition means that we don't need to introduce boundary degrees of freedom to compensate gauge anomalies as the gauge symmetry is broken at the boundary. On the other hand, we are forced to contend with non-perturbative corrections by boundary monopole operators. 

Many aspects of these boundary monopole operators are understood, \cf\, \cite[Section 7]{CostelloDimofteGaiotto-boundary}, but an explicit, widely applicable description is still being developed. Nonetheless, it is straightforward to reproduce the perturbative analysis of \cite{CostelloGaiotto} using the twisted theory presented above. Explicitly, the only local operators that survive at the boundary are built from the boundary values of the lowest components $B, \phi, \psi_X, \psi_Y$ of the bulk twisted superfields $\bB, \bPhi, \bPsi_{\bX}, \bPsi_{\bY}$. Moreover, their $\oz$ dependence is trivialized in cohomology. One finds that the non-trivial OPEs of these operators are the following%
\footnote{Viewing the $B$-twist deformation terms $\int \bY \pd \bX - \bLambda \pd \bA$ as Chern-Simons terms, we see that the mixed level for $B, \phi$ is $-1$. We have also rescaled $\psi_Y, \phi$ by a factor of $2\pi$. We will see this negative sign and rescaling again in Section \ref{sec:SQEDrev}.}:
\begin{gather}
	\label{eq:opeSYMB}
	B_a(z) B_b(w) \sim  \frac{(T_V - 2h_G)\kappa_{ab}}{(z-w)^2} + \frac{f^c{}_{ab}}{z-w} B_c(w) \qquad \psi_Y{}^n(z) \psi_X{}_m(w) \sim \frac{\delta^n{}_m}{(z-w)^2} + \frac{(\tau_a)^n{}_m}{z-w} \phi^a(w) \nonumber \\
	B_a(z) \phi^b(w) \sim  -\frac{\delta^b{}_a}{(z-w)^2} + \frac{f^b{}_{ca}}{z-w}\phi^c(w)\\
	B_a(z) \psi_Y{}^n(w) \sim -\frac{(\tau_a)^n{}_m}{z-w}\psi_Y{}^m(w) \qquad B_a(z) \psi_X{}_n(w) \sim \frac{ (\tau_a)^m{}_n}{z-w}\psi_X{}_m(w) \nonumber
\end{gather}
This can be interpreted as an affine current algebra based on the above Lie superalgebra, with central extension determined by the anomaly for the boundary $G$ symmetry, the pairing of $\mathfrak{g}$ and $\mathfrak{g}^*$, and the symplectic form on $T^*V$.

\section{VOAs for Twisted Chern-Simons-Yang-Mills Theories}
\label{sec:CSYM}
We now turn to the main theories of interest. The Chern-Simons-matter theories we consider can be realized by gauging hypermultiplet flavor symmetries of the $\CN=4$ Yang-Mills gauge theories of Section \ref{sec:twistedVOAs} using $\CN=2$ vector multiplets with an $\CN=2$ Chern-Simons term, as opposed to the usual gauging with $\CN=4$ vector multiplets. The resulting theory has at least $\CN=2$ supersymmetry and hence is guaranteed to admit a holomorphic-topological twist. 

Just as in Sections \ref{sec:SYMA} and \ref{sec:SYMB}, we seek topological deformations of the $HT$ twist for these Chern-Simons-matter theories. If the resulting Chern-Simons-matter theory is itself an $\CN=4$ theory, such deformations can arise from deforming to the theory's $A$ and $B$ twists. Gaiotto and Witten described a class of such $\CN=4$ Chern-Simons-matter theories constructed as above, \ie\, by gauging a flavor symmetry of a system of $\CN=4$ hypermultiplets using an $\CN=2$ vector multiplet with Chern-Simons term (together with a superpotential) \cite{GaiottoWitten-Janus}. In most cases, gauging hypermultiplets with $\CN=2$ Chern-Simons fields yields a theory with only $\CN=3$ supersymmetry \cite{GaiottoYin, dMFOME}. This is enhanced to $\CN=4$ supersymmetry when the moment map for the corresponding symmetry satisfies a ``fundamental identity": to gauge an $H_c$ flavor symmetry of hypermultiplets transforming in a representation $R$, the complex moment map $\nu_\C: R \to \mathfrak{h}^*$ must be isotropic with respect to the $H_c$-invariant, non-degenerate, bilinear form $\Tr$ (whose inverse, also denoted $\Tr$, appears in the Chern-Simons term)
\be
\label{eq:fundid}
	\Tr(\nu_\C{}^2) = 0
\ee
for any choice of complex structure on $R$. This can be relaxed to simply requiring that $\Tr(\nu_\C^2)$ is constant (although the constant can depend on the choice of complex structure).%
\footnote{This weakened constraint has an alternative interpretation as extending the (complexified) gauge group $H$ to a supergroup $\widehat{H}$ by odd directions parameterized by $R$, correspondingly $\mathfrak{h}$ extends to the Lie superalgebra $\widehat{\mathfrak{h}} \cong \mathfrak{h} \oplus \Pi R$ with brackets in Eq. \eqref{eq:CSBliealg}. The fundamental identity arises as the Jacobi identity for three odd generators. There are various perspectives on this supersymmetry enhancement; see, \eg\,, \cite{dMFOME, N=34}.}

\begin{figure}[h!]
	\centering
	\begin{tikzpicture}[scale=1]
		\begin{scope}[auto, every node/.style={inner sep=1}, node distance=1cm]
			\node (v3) {};
			\node[right=of v3] (v4) {};
		\end{scope}
		\begin{scope}[auto, every node/.style={minimum size=3em,inner sep=1},node distance=0.5cm]
			\node[draw, circle, left=of v3] (v2) {$n-1$};
			\node[draw, left=of v2] (v1) {$n$};
			\node[draw, circle, right=of v4] (v5) {2};
			\node[draw, circle, right=of v5] (v6) {1};
		\end{scope}
		\draw(v1)--(v2);
		\draw(v2)--(v3);
		\draw[loosely dotted, line width = 1pt] (v3)--(v4);
		\draw(v4)--(v5);
		\draw(v5)--(v6);
	\end{tikzpicture}
	\caption{Quiver describing the 3d $\CN=4$ gauge theory that flows to $T[SU(n)]$ in the IR.}
	\label{fig:linquiver}
\end{figure}
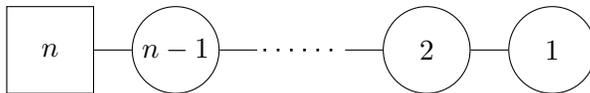

This construction can be extended to non-linear, hyperk\"ahler target spaces, so long as the corresponding fundamental identity holds. A particularly important class of examples comes from Higgs branches of more traditional $\CN=4$ Yang-Mills theories. The Higgs branch always admits a tri-Hamiltonian action by (any subgroup of) the (compact) Higgs-branch flavor symmetry $H_c$. If the $H_c$ action on the Higgs branch satisfies the fundamental identity, then it is possible to gauge the symmetry with Chern-Simons fields in a way that the resulting theory flows to an $\CN=4$ theory in the IR. For example, the work \cite{KapustinSaulina-CSRW} considers a Chern-Simons gauging of the $SU(n)$ flavor symmetry of hypermultiplets valued in the cotangent bundle to the flag variety $T^*\Fl_n$. This cotangent bundle is the (resolved) Higgs branch of the 3d $\CN=4$ theory $T[SU(n)]$ living on the $S$-duality interface of 4d $\CN=4$ $SU(n)$ super Yang-Mills \cite{GaiottoWitten-Sduality}; $T[SU(n)]$ can be realized as the IR limit of the $\CN=4$ gauged linear $\sigma$-model given by the quiver in Figure \ref{fig:linquiver}.

Generally, the UV description of such a gauging does not admit $\CN=4$ supersymmetry as the fundamental identity holds only on the Higgs branch, \ie\, only after the hyperk\"ahler moment maps for action of the Yang-Mills gauge group $G_c$ are set to the corresponding FI parameters. Nonetheless, we find that they admit $A$ and $B$ twist topological deformations once we pass to a description in terms of twisted superfields. In other words, the obstruction to having $\CN=4$ supersymmetry in the UV belongs to the kernel of the quasi-isomorphism of passing to $HT$-twisted superfields. The $B$-twisted theories we describe in Section \ref{sec:CSYMB} can be thought of as gauged linear $\sigma$-model versions of the Chern-Simons-Rozansky-Witten theories introduced in \cite{KapustinSaulina-CSRW}, and agrees with the AKSZ theory of \cite{KQZ} when there are no Yang-Mills gauge fields.

Somewhat more surprisingly, we find that there is a topological ``$A$ twist" regardless of the $\CN=4$ theory we are gauging with $\CN=2$ vector multiplets.%
\footnote{We call this the ``$A$ twist" because it agrees with the $A$ twist in the absence of Chern-Simons fields. Importantly, a general hypermultiplet target will \emph{not} give an $\CN=4$ theory, nor is such a theory expected to flow to an $\CN=4$ theory in the IR. Thus, we should not expect a ``$B$ twist" unless the fundamental identity is satisfied on the Higgs branch. We will see this in Section \ref{sec:CSYMB}.}
In particular, using Chern-Simons fields to gauge Higgs-branch flavor symmetries of standard $\CN=4$ theories is compatible with the $A$ twist deformation described in Section \ref{sec:SYMA} independent of the fundamental identity. Moreover, based on the argument in \cite[Section 2.3]{twistedN=4}, the resulting theory is perturbatively topological.%
\footnote{As described in \emph{loc. cit.}, the fact that the theories described in this paper are perturbatively topological follows from checking that the classical theory is topological via a homotopy trivialization of holomorphic translations (and more general holomorphic vector fields), \cf\, \cite[Def 5.2]{EGW}. By the results of \cite{GwilliamWilliams}, the corresponding trivialization doesn't suffer from a perturbative anomaly, hence the theory continues to be topological in perturbation theory. See \cite{EGW} for a related discussion in the context of deforming holomorphic twists of 4d $\CN=4$ super Yang-Mills.

We note that non-perturbative obstructions are entirely possible, and can occur in related analyses in 2d $\CN=(0,2)$ theories \cite{SilversteinWitten2}. We do not expect non-perturbative effects to alter the topological nature of the $A$- and $B$-twisted Gaiotto-Witten theories described in this paper as they should correspond to twists by an honest (non-anomalous) action of $\CN=4$ supersymmetry, but our ``$A$-twisted" theories are not as robust. We leave a full, non-perturbative verification of the topological nature of these theories to future work.} %
The existence of these topological theories for general Higgs branch flavor symmetry is totally unexpected. Roughly speaking, the $A$ twist of the to-be-gauged 3d $\CN=4$ theory trivializes the fundamental identity for Higgs-branch flavor symmetries, thereby allowing the symmetry to be gauged with Chern-Simons fields in a way compatible with the twist.

In Section \ref{sec:bdyCSYMA} we discuss boundary conditions for these ``$A$-twisted" theories that furnish interesting boundary VOAs realizing cosets of the VOAs of \cite{CostelloGaiotto} by affine subalgebras. As an example, in Section \ref{sec:gaugedhyper} we identify a boundary VOA for $U(1)_1$ coupled to a single charge 1 hypermultiplet, finding Kausch's $p=2$ singlet VOA $\mathfrak{M}(2)$ \cite{Kausch}, also called Zamolodchikov's $\mathcal{W}_3$ algebra with $c=-2$ \cite{ZamoloWalgs}, realized as the coset of symplectic bosons by its Heisenberg subalgebra. This log-VOA has a tensor category of modules with an (uncountably) infinite number of simple modules (as well as reducible but indecomposable modules) and whose modular data has been well studied, see \eg\, \cite{CRlogCFT} and references therein. A straightforward half-index computation suggests that Wilson lines in the 3d bulk should be identified with the atypical simple modules of the singlet $\mathfrak{M}(2)$. It would be interesting to identify the bulk lines corresponding to the remaining simples.

More generally, we expect these ``$A$-twisted" TQFTs have highly non-trivial algebras of local operators and, correspondingly, their categories of line operators should be non-semisimple tensor categories realizing an admixture of Wilson lines in Chern-Simons theories and the vortex line operators of super Yang-Mills theories. In the recent paper \cite{CDGG}, we pursue this idea in more detail for the 3d $\CN=4$ theory obtained from using Chern-Simons fields to gauge the $SU(n)$ Higgs-branch flavor symmetry of $T[SU(n)]$ and relate the corresponding category of line operators to the representation theory of the quantum group $U_q(\mathfrak{sl}(n))$ at even roots of unity $q^{2k} = 1$.

\subsection{``$A$ twist"}
\label{sec:CSYMA}

First consider the case where we have hypermultiplets valued in a (complex) symplectic vector space $T^*U$, where $U$ is an $H$-representation such that the $H$-action on $T^*U$ satisfies the fundamental identity. Equivalently, we could require that $T^*U$ constitutes the odd subspace of a Lie superalgebra $\widehat{\mathfrak{h}}$ whose even subalgebra is $\mathfrak{h}$. This $\CN=4$ Chern-Simons-matter theory is a special case of the $\CN=3$ Chern-Simons-matter theory based on general hypermultiplet target and can be easily written in terms of $\CN = 2$ superfields. It contains an $\CN = 2$ vector multiplet and chiral/anti-chiral multiplets coming from the hypermultiplets, with a superpotential roughly given by $\Tr(\nu_\C{}^2)$ \cite{GaiottoYin}.\footnote{In the $\CN=3$ theory, the adjoint chiral multiplet $\Phi$ in the $\CN = 4$ vector multiplet serves as an auxiliary field and only appears in the superpotential, which is roughly given by $-\tfrac{k}{8\pi} \Tr(\Phi^2) + \nu_\C \Phi$. Plugging in the equations of motion for $\Phi$ gives a superpotential of the form $\tfrac{2\pi}{k}\Tr(\nu_\C{}^2)$. Note that this superpotential is not invariant under $U(1)_F$ unless it is a constant, indicating the enhancement from $\CN = 3$ to $\CN = 4$ when $\tfrac{2\pi}{k}\Tr(\nu_\C{}^2) = 0$.} Thus, when written in $\CN=2$ language, the superpotential for the Gaiotto-Witten $\CN=4$ Chern-Simons theories with linear hypermultiplets is at worst a constant, and must vanish if we wish to preserve $U(1)_A \times U(1)_B$. The action for this $HT$-twisted Gaiotto-Witten theory (for linear target) is thus given by
\be
\label{eq:twistedactionGW}
	S = \int \bB F'(\bA)  + \tfrac{k}{4\pi} \Tr(\bA \pd \bA) + \bPsi_{\bX} \diff'_{\bA} \bX + \bPsi_{\bY} \diff'_{\bA} \bY\,.
\ee
and the action of $Q_{HT}$ is then given by
\be
\begin{aligned}
	\label{eq:QGW}
	Q_{HT} \bA & = F'(\bA) \qquad & Q_{HT} \bB & = \diff'_{\bA} \bB - \bnu + \tfrac{k}{2\pi}\pd \bA \\
	Q_{HT} \bX & = \diff'_{\bA} \bX \qquad & Q_{HT} \bPsi_{\bX} & = \diff'_{\bA} \bPsi_{\bX}\\
	Q_{HT} \bY & = \diff'_{\bA} \bY \qquad & Q_{HT} \bPsi_{\bY} & = \diff'_{\bA} \bPsi_{\bY}\\
\end{aligned}\,.
\ee
It is worth noting that the above $HT$-twisted action doesn't need the hypermultiplet representation to satisfy the fundamental identity. Indeed, the superpotential $W \sim \Tr(\nu_\C{}^2)$ is required for $\CN \geq 3$ supersymmetry, but there is an $HT$ twist even in its absence.

The transformations of the hypermultiplets \cite{KapustinSaulina-CSRW, GaiottoWitten-Janus} suggest that we should again include the deformation $-\int \bPsi_{\bX} \bPsi_{\bY}$, \cf\, the deformation in Section \ref{sec:SYMA}. Similarly, the transformation of the gauge field contains the scalars in the anti-chiral multiplet, which suggests $\delta_A \bA = 0$. In particular, we consider a deformed action of the form 
\be
\label{eq:AtwistedactionGGW}
S_{A} = S - \int \bPsi_{\bX} \bPsi_{\bY}\,.
\ee
From this, we find that the action of $Q_A$ is given by
\be
\begin{aligned}
	\label{eq:QAGW}
	Q_A \bA & = F'(\bA) \qquad & Q_A \bB & = \diff'_{\bA} \bB - \bnu + \tfrac{k}{2\pi}\pd \bA \\
	Q_A \bX & = \diff'_{\bA} \bX - \bPsi_{\bY} \qquad & Q_A \bPsi_{\bX} & = \diff'_{\bA} \bPsi_{\bX}\\
	Q_A \bY & = \diff'_{\bA} \bY + \bPsi_{\bX} \qquad & Q_A \bPsi_{\bY} & = \diff'_{\bA} \bPsi_{\bY}\\
\end{aligned}.
\ee
Note that the nilpotence of $Q_A$ does \emph{not} depend on the fundamental identity for the hypermultiplets. 

The equations of motion for the fermions $\bPsi_{\bX}, \bPsi_{\bY}$ are purely algebraic, and if we were to substitute those solutions into the action, we find
\be
	S'_A = \int \widehat{\bB} F'(\bA) + \tfrac{k}{4\pi}\Tr\big(\bA \pd \bA\big)\,,
\ee
where $\widehat{\bB} = \bB + \bnu_\C$, up to boundary terms.%
\footnote{There should also be a 1-loop correction to the level $k$ proportional to the quadratic index $T_U$ coming from integrating out the hypermultiplet fermions.} %
We see that the equations of motion for $(\bA, \widehat{\bB})$
\be
\begin{aligned}
	Q_A \bA & = F'(\bA) \qquad & Q_A \widehat{\bB} & = \diff'_{\bA} \widehat{\bB} + \tfrac{k}{2\pi}\pd \bA
\end{aligned}
\ee
are those of usual $\mathfrak{h}_k$ Chern-Simons theory, in agreement with the results of \cite{KLLtop}.

It is straightforward to include $\CN=4$ vector multiplets in this ``$A$-twisted" theory. The (complexified) gauge group takes the form $G \times H$, where the Yang-Mills gauge fields are valued in $\mathfrak{g}$ and the Chern-Simons gauge fields are valued in $\mathfrak{h}$. We will denote the vector multiplet twisted superfields as $(\bA = \bA_{\rm YM} \oplus \bA_{\rm CS}, \bB = \bB_{\rm YM} \oplus \bB_{\rm CS})$, and denote the fields coming from the adjoint chiral in the $\CN=4$ vector multiplet as $(\bPhi, \bLambda)$. The hypermultiplets will be denoted as above and we assume that they transform in a representation $R = \bigoplus T^*(V_i \times U_i)$, where $V_i$ (resp. $U_i$) is a representation of $G$ (resp. $H$). The ``$A$-twisted" action is given by
\be
\label{eq:AtwistedactionCSYM}
\begin{aligned}
	S_A & = \int \bB_{\rm YM} F'(\bA_{\rm YM}) + \bLambda \diff'_{\bA_{\rm YM}} \bPhi + \bB_{\rm CS} F'(\bA_{\rm CS}) +\tfrac{k}{4\pi} \Tr(\bA_{\rm CS} \pd \bA_{\rm CS})\\
	& \qquad + \bPsi_{\bX} \diff'_{\bA} \bX + \bPsi_{\bY} \diff'_{\bA} \bY + \bY \bPhi \bX + \bB_{\rm YM} \bPhi - \bPsi_{\bX} \bPsi_{\bY}
\end{aligned},
\ee
which yields an action of $Q_A$ given by
\be
\label{eq:QACSYM}
\begin{aligned}
	Q_A \bA_{\rm CS} & = F'(\bA_{\rm CS}) \qquad & Q_A \bB_{\rm CS} & = \diff'_{\bA_{\rm CS}} \bB_{\rm CS} - \bnu + \tfrac{k}{2\pi}\pd \bA_{\rm CS}\\
	Q_A \bA_{\rm YM} & = F'(\bA_{\rm YM}) + \bPhi \qquad & Q_A \bB_{\rm YM} & = \diff'_{\bA_{\rm YM}} \bB_{\rm YM} - \bmu\\
	Q_A \bPhi & = \diff'_{\bA_{\rm YM}} \bPhi & Q_A \bLambda & = \diff'_{\bA_{\rm YM}} \bLambda + \bmu_\C + \bB_{\rm YM}\\
	Q_A \bX & = \diff'_{\bA} \bX - \bPsi_{\bY} \qquad & Q_A \bPsi_{\bX} & = \diff'_{\bA} \bPsi_{\bX} + \bY \bPhi\\
	Q_A \bY & = \diff'_{\bA} \bY + \bPsi_{\bX} \qquad & Q_A \bPsi_{\bY} & = \diff'_{\bA} \bPsi_{\bY} + \bPhi \bX\\
\end{aligned}.
\ee
Again, the nilpotence of $Q_A$ is independent of the fundamental identity so this theory is consistent for arbitrary choice of $H,G,R$. The equations of motion for the fermions $\bPsi_{\bX}, \bPsi_{\bY}$ and the bosons $\bB_{\rm YM}, \bPhi$ are purely algebraic and plugging them into Eq. \eqref{eq:AtwistedactionCSYM} again yields a Chern-Simons action with fields $(\widehat{\bB}_{\rm CS} = \bB_{\rm CS} + \bnu_{\C}, \bA_{\rm CS})$ as above.

As described in \cite[Sectin 2.3]{twistedN=4}, we can verify that this theory is perturbatively topological by identifying a (classical) local operator $\bS$ such that $Q_A \bS = \diff \bS + \bT$ for $\bT$ the (classical) stress tensor superfield. In this ``$A$-twisted" theory, we find that the stress tensor is given by
\be
	\bT = \bB_{\rm CS} \pd_z \bA_{\rm CS} - \bB_{\rm YM} \pd_z \bA_{\rm YM} + \bLambda \pd_z \bPhi + \tfrac{1}{2}\big(\bPsi_{\bX} \pd_z \bX + \bPsi_{\bY} \pd_z \bY - \bX \pd_z \bPsi_{\bX} - \bY \pd_z \bPsi_{\bY}\big)
\ee
from which it is easy to check that 
\be
	\bS = -\tfrac{\pi}{k} \Tr\widehat{\bB}_{\rm CS}\iota_{\pd_z} \widehat{\bB}_{\rm CS} - \bLambda \pd_z \bA_{\rm YM} + \tfrac{1}{2}(\bY \pd_z \bX - \bX \pd_z \bY)
\ee
is the desired homotopy trivialization.

\subsubsection{Boundary algebra for the ``$A$ twist"}
\label{sec:bdyCSYMA}
We now describe a boundary algebra for the above ``$A$-twisted" theories. First consider the case of hypermultiplets transforming in a symplectic representation $R = T^*U$ of a gauge group $H$, where we gauge $H$ with an $\CN=2$ vector multiplet with Chern-Simons level $k$. In terms of twisted superfields, the boundary conditions of interest are the following.
\begin{itemize}
	\item Neumann boundary conditions for the vector multiplets. ($\bB|_{\pd} = \bmu_\pd$)
	\item Neumann boundary conditions for the hypermultiplets. ($\bPsi_{\bX}|_{\pd}, \bPsi_{\bY}|_{\pd} = 0$)
	\item Extra (chiral) boundary degrees of freedom, with a $H$ flavor symmetry, to cancel any gauge anomalies.
\end{itemize}
These choices ensure that there are no boundary monopole operators and thus it suffices to consider a perturbative analysis. The gauge anomaly for the gauge group $H$ can be canceled just as before; we assume the level $k$ is sufficiently negative%
\footnote{As mentioned above, we could instead work with sufficiently positive levels if we exchange left and right boundary conditions.} %
so that the anomaly can be canceled with boundary Fermi multiplets transforming in a representation $N$ with quadratic index satisfying $T_N = T_U - (k + h_H)$. We will denote the boundary fermions by $\gamma, \wt{\gamma}$ as before.

The perturbative analysis is nearly identical to that performed in Section \ref{sec:SYMA}: the only local operators that survive in cohomology are built from the boundary values of the bosons at the bottom of the hypermultiplet $X,Y$ (which furnish a symplectic boson VOA $\Sb[T^*U]$), the boundary fermions $\gamma, \wt{\gamma}$ (which furnish a complex fermion VOA $\Ff[N]$), and the ghost $c$ (which has trivial OPEs with everything). In addition, there is a non-trivial action of the supercharge $Q_A$ given by
\begin{gather}
	Q_A c = c^2\\
	\begin{aligned}
		Q_A X & = c \cdot X \qquad & Q_A Y & = c \cdot Y\\
		Q_A \gamma & = c \cdot \gamma \qquad & Q_A \widetilde{\gamma} & = c \cdot \widetilde{\gamma}\\
	\end{aligned}\,.
\end{gather}
As described in, \eg\,, \cite[Section 6.2]{CostelloDimofteGaiotto-boundary}, we should not include the zero-mode of the $c$ ghost and instead take invariants with respect to constant gauge transformations ($H$ invariants) by hand and not with a ghost; this is simply the statement that we must take \emph{derived} invariants with respect to the group $H[\![z]\!]$ of holomorphic gauge transformations
\be
	\CV_A[R, H_k; N] = H^\bullet\big(\langle\!\langle X, Y, \gamma, \wt{\gamma}, \pd c \rangle\!\rangle^H, Q_A\big) =: \big(\Sb[T^*U] \times \Ff[N]\big)^{H[\![z]\!]}\,.
\ee

Importantly, the action of $H[\![z]\!]$ on $\Sb[T^*U] \times \Ff[N]$ is internal, \ie\, there are affine currents $J_{\mathfrak{h}}$ at level $- (k + h_H)$ whose OPEs generate the $H[\![z]\!]$ action. Thus, the differential $Q_A$ acts on an operators $\CO$ that is independent of $c$ (and its derivatives) as
\be
\label{eq:QACSbdy}
	Q_A \CO(z) = -\sum \limits_{\ell \geq 0} \frac{\pd^\ell c_(z)}{\ell!}\oint_{z} \frac{\diff w}{2\pi i} (z-w)^\ell J_{\mathfrak{h}}(w) \CO(z).
\ee
The form of the action of $Q_A$ suggests that the boundary algebra can be described simply as a coset. More precisely, so long as the derived $H[\![z]\!]$ invariants are concentrated in degree 0, we expect that the boundary algebra is described by the coset
\be
	\CV_A[R, H_k; N] = \frac{\Sb[R] \times \Ff[N]}{H_{-(k+h_H)}}\,.
\ee
In such a situation, the boundary algebra has a holomorphic stress tensor $T(z)$ (the coset stress tensor, with central charge given by the coset central charge). It is important to note that the boundary algebra cannot have a holomorphic stress tensor if the derived $H[\![z]\!]$ invariants have support in non-zero cohomological degrees. This can be seen by noting that $c$ has trivial OPEs with all operators, thus any putative stress tensor cannot have the correct OPE with operators built from $c$ or its derivatives. We illustrate an example that satisfies these criteria in the next section, namely we consider a single hypermultiplet $U = \C$ and with gauge group $H = U(1)$ at level $k = 1$. The existence of such a boundary stress tensor strongly suggesting the bulk theory is topological beyond perturbation theory.

Now consider the general case with both $\CN=4$ vector multiplets and Chern-Simons vector multiplets. We assume the hypermultiplets take values in a representation of the form $R = \bigoplus_i T^*(V_i \times U_i)$ for $U_i$ (resp. $V_i$) a complex representation of $H$ (resp. $G$). In the absence of Chern-Simons gauge fields ($H = 1$), we recover the boundary algebras discussed in Section 4 of \cite{CostelloGaiotto} and Section \ref{sec:SYMA} above. In terms of twisted superfields, the boundary conditions of interest are the following.
\begin{itemize}
	\item Neumann boundary conditions for the vector multiplets. ($\bB|_{\pd} = \bmu_\pd$)
	\item Dirichlet boundary conditions for the adjoint chiral multiplet. ($\bPhi|_{\pd} = 0$)
	\item Neumann boundary conditions for the hypermultiplets. ($\bPsi_{\bX}|_{\pd}, \bPsi_{\bY}|_{\pd} = 0$)
	\item Extra (chiral) boundary degrees of freedom, with a $G \times H$ flavor symmetry, to cancel any gauge anomalies.
\end{itemize}
These choices ensure that the superpotential $\bY \bPhi \bX + \bB_{\rm YM}\bPhi - \bPsi_{\bX} \bPsi_{\bY}$ vanishes as the boundary. Moreover, the Neumann boundary conditions on the vector multiplets ensure that there are no boundary monopole operators and thus it suffices to consider a perturbative analysis. The gauge anomaly for the Yang-Mills gauge group $G$ can be canceled just as before; we assume that $\sum (\dim U_i) T_{V_i} \geq 2 h_G$ and cancel the anomaly with free fermions transforming in a representation $M$ such that $T_{M} = \sum (\dim U_i) T_{V_i} - 2 h_G$. We also assume that the anomaly for the Chern-Simons gauge group $H$ can be canceled with boundary Fermi multiplets transforming in a representation $N$ with $T_N = \sum (\dim V_i) T_{U_i} - (k + h_H)$, which can be done for sufficiently negative $k$. We will collectively denote the boundary fermions by $\gamma = \gamma_M \oplus \gamma_N, \wt{\gamma} = \wt{\gamma}_M \oplus \wt{\gamma}_N$.

The perturbative analysis is again nearly identical to that performed in Section \ref{sec:SYMA}. The boundary algebra is generated by $G \times H$ invariants of the lowest components of the bulk and boundary matter fields $\lambda, X, Y, \gamma, \wt{\gamma}$ together with the (derivatives of the) ghosts $c_{\rm YM}, c_{\rm CS}$ with non-trivial OPEs given by Eq. \eqref{eq:SYMAope} (with $c$ replaced by $c_{\rm YM}$)
\be
c^a_{\rm YM}(z) \lambda_b(w) \sim \frac{\delta^a{}_b}{z-w} \qquad X^n(z) Y_m(w) \sim \frac{\delta^n{}_m}{z-w} \qquad \gamma^p(z) \wt{\gamma}_q(w) \sim \frac{\delta^p{}_q}{z-w}
\ee
and subject to the differential
\begin{gather}
	Q_A c_{\rm CS} = c_{\rm CS}^2\\
	\begin{aligned}
		Q_A c_{\rm YM} & = c_{\rm YM}^2 \qquad & Q_A \lambda & = J_{\mathfrak{g}}\\
		Q_A X & = c \cdot X \qquad & Q_A Y & = c \cdot Y\\
		Q_A \gamma & = c \cdot \gamma \qquad & Q_A \widetilde{\gamma} & = c \cdot \widetilde{\gamma}\\
	\end{aligned}
\end{gather}
where $J_{\mathfrak{g}}$ is the current generating the affine $\mathfrak{g}$ action
\be
\label{eq:CSYMbdyVOA}
	\CV_{A}[R, G \times H_k; M \oplus N] = H^\bullet \big(\langle\!\langle X, Y, \gamma, \wt{\gamma}, \lambda, \pd c_{\rm YM}, \pd c_{\rm CS} \rangle\!\rangle^{G \times H}, Q_A\big)\,.
\ee

We expect that the resulting algebra can be computed with a spectral sequence as follows. We first ignore the Chern-Simons fields and the boundary fermions used to cancel the associated anomalies, take $G$-invariants, and work with the differential coming from the super Yang-Mills theory and the $M$-valued fermions $\gamma_M, \wt{\gamma}_M$. This process once again yields the $A$ twist boundary VOA $\CV_A[G, R; M]$ of \cite{CostelloGaiotto}: the $G$-BRST reduction of $R$ valued symplectic bosons and $M$ valued complex fermions
\be
	\CV_A[G, R; M] = \big(\Sb[R] \times \Ff[M]\big)/\!/G\,.
\ee
We then take $H$-invariants, followed by cohomology with respect to the second differential implementing derived $H[\![z]\!]$-invariants of the product of this $G$-BRST reduction and the $N$-valued fermions. So long as the $G$-BRST reduction is concentrated in degree 0, which we expect to hold in most examples of interest, the spectral sequence degenerates after this page.

Just as above, the action of $H[\![z]\!]$ on $\CV_A[G, R; M] \times \Ff[N]$ is internal. Thus, so long as the derived $H[\![z]\!]$ invariants are concentrated in degree 0, we expect that the boundary algebra is described by the coset
\be
\CV_A[R, G \times H_k; M, N] = \frac{\CV_A[G, R; M] \times \Ff[N]}{H_{-k-h_H}}\,.
\ee
In particular, the boundary algebra has a holomorphic stress tensor $T(z)$ (with central charge given by the coset central charge), strongly suggesting the bulk theory is topological.

It is important to note that the boundary algebra cannot have a holomorphic stress tensor if the derived $H[\![z]\!]$-invariants have support in non-zero cohomological degrees. This can be seen by noting that $c_{\rm CS}$ has trivial OPEs with all operators, thus any putative stress tensor cannot have the correct OPE with operators built from $c_{\rm CS}$ or its derivatives.

\subsubsection{Example: $U(1)_1$ gauged hypermultiplet}
\label{sec:gaugedhyper}
In this section, we consider an example ``$A$-twisted" theory obtained by coupling $U(1)_1$ Chern-Simons gauge fields to a single $A$-twisted hypermultiplet of charge $1$.%
\footnote{The work \cite{GKLSYrank0} predicted that this theory actually flows to an $\CN=5$ SCFT in the IR. This enhancement is not expected to be the origin of the topological described in this section: the ``$A$ twist" we describe is present for any $k$, whereas the enhancement to $\CN=5$ only happens for $|k|=1$.} 
See Fig. \ref{fig:U11plushyper}. Since the Chern-Simons level is $1$, the boundary $U(1)$ gauge symmetry is non-anomalous and we do not need to introduce any boundary degrees of freedom. The boundary VOA is simply given by the derived $\C^\times[\![z]\!]$-invariants of the symplectic boson VOA. 

Concretely, we have the ghost $c = c_{\rm CS}$ and the symplectic bosons $X,Y$, with the only non-trivial OPE given by
\be
X(z) Y(w) \sim \frac{1}{z-w}\,.
\ee
There is a differential given by
\be
Q_A c = 0 \qquad Q_A X = c X \qquad Q_A Y = - c Y\,.
\ee
Note that the action of $Q_A$ on operator built from the symplectic bosons $X,Y$ can be realized as in Eq. \eqref{eq:QACSbdy} with the current $J_{\mathfrak{gl}(1)} = J = \norm{Y X}$. Importantly, any operator with non-trivial dressing by derivatives of $c$ must be $Q_A$-exact. For example,
\be
\begin{aligned}
	Q_A\norm{Y X}(z) & = \lim\limits_{w \to z} Q_A\bigg(Y(w) X(z) - \frac{1}{z-w}\bigg)\\
	& = \lim\limits_{w \to z} (-c(w) Y_m(w)) X^n(z) + Y_m(w)(c(z) X^a(z))\\
	& = \lim\limits_{w\to z} \frac{c(z) - c(w)}{z-w} = \pd c(z)
\end{aligned}\,,
\ee
\cf\, Section 6.4 of \cite{CostelloDimofteGaiotto-boundary}. Moreover, operators that are independent of the ghosts cannot be $Q_A$-exact for degree reasons. We conclude that the boundary algebra is concentrated in degree 0, and agrees with the coset of the symplectic bosons $X,Y$ by the current subalgebra generated by $J$:
\be
	\CV_\pd[U(1)_1 + \textrm{hypermultiplet}] = \frac{\Sb}{U(1)_{-1}}\,.
\ee

This is a well studied coset, and results in Kausch's $p=2$ singlet VOA $\mathfrak{M}(2)$ \cite{Kausch,WangWalgs}, also called Zamolodchikov's $\mathcal{W}_3$ algebra with $c=-2$ \cite{ZamoloWalgs}. Indeed, the half-index counting boundary local operators is easily computed to be
\be
\label{eq:halfind}
I\!\!I(q) = (q)_\infty \oint \frac{\diff s}{2\pi i s}\frac{1}{(q^{\tfrac{1}{2}}s;q)_\infty (q^{\tfrac{1}{2}}s^{-1}; q)_\infty} = \sum\limits_{j \geq 0}\frac{(-1)^j q^{\tfrac{j(j+1)}{2}}}{(q)_\infty}\,,
\ee
where we have used the standard notation
\be
(x;q)_\infty = \prod_{n \geq 0}(1-x q^n) \qquad (q)_\infty = (q;q)_\infty
\ee
for the $q$-Pochhammer symbol, exactly matching the vacuum character for the singlet VOA \cite{CRlogCFT} up to an overall factor of $q$:
\be
	\chi({\rm vacuum}) = q^{-\tfrac{c}{24}}I\!\!I(q)\,.
\ee

The singlet VOA $\mathfrak{M}(2)$ is one of the simplest examples of a logarithmic conformal field theory. Its representation theory and modular properties have been thoroughly studied; there are an uncountable number of irreducible modules, as well as reducible but indecomposable modules. For more details, see \eg\, \cite{CRlogCFT} and references therein. We expect that the atypical, irreducible modules $\CM_{n}$ correspond to bulk Wilson lines ending on the boundary: the half-index in the presence of the Wilson line $\W_{-n}$ of charge $-n$ yields
\be
\label{eq:halfindWilson}
	I\!\!I(q;\W_{-n}) = (q)_\infty \oint \frac{\diff s}{2\pi i s} s^{-n} (s q^{\tfrac{1}{2}};q)_\infty (s^{-1}q^{\tfrac{1}{2}}; q)_\infty = \sum\limits_{j \geq 0}\frac{(-1)^j q^{\tfrac{j(j+1)+n(2j+1)}{2}}}{(q)_\infty}\,,
\ee
which exactly matches the character of the atypical module $\CM_{n}$, up to an overall factor of $q$
\be
	\chi(\CM_n) = q^{\tfrac{n^2}{2} - \tfrac{c}{24}} I\!\!I(q; \W_{-n})\,.
\ee
It would be interesting to see how the typical modules $\CF_\mu$ are realized in terms of our bulk ``$A$-twisted" Chern-Simons theory.

Based on standard manipulations, the algebra of local operators in the bulk TQFT can be extracted from the category of line operators as the self-extensions of the trivial line operator. In terms of modules for the boundary algebra $\mathfrak{M}(2)$, this statement translates to the self-extensions of the vacuum module within a suitable category of modules. Thankfully, this is known: there are only trivial self-extensions%
\footnote{The absence of self-extensions depends on what category of modules one works with, see \eg\, \cite[Section 9]{CostelloCreutzigGaiotto}. In particular, to obtain this result we must work in a category containing all finite length modules, not just ones generated by simple objects. More generally, allowing all finite length modules seems to be the correct choice if one is interested in identifying VOA modules with bulk line operators in topologically twisted 3d theories.} %
of the vacuum module \cite{milasintertwiners, AMintertwiners, CMRunrolled}, from which it follows that the bulk algebra of local operators should be trivial! Indeed, this agrees with the superconformal index of the bulk theory \cite{GKLSYrank0} as well as a straightforward generalization of the non-perturbative analysis described in \cite{Zeng}.%
\footnote{To sketch how this latter point is verified, note that the bulk Chern-Simons level gives electric charge to bulk monopoles and so they must be dressed by the charged fields $X, Y, \psi_X, \psi_Y$, but all such dressings are $Q_A$-exact if they are $Q_A$-closed.} %

The supersymmetric gauge theory underlying this ``$A$-twisted" theory is an example studied in \cite{GKLSYrank0}. The authors of \emph{loc. cit.} compute various limits of superconformal indices of this theory, finding a trivial answer (in agreement with the above conclusion of a trivial algebra of local operators). They further extract modular data for two TQFTs from three-sphere partition functions. It would be interesting to try to understand how their analysis can be understood from the present perspective.

We end this example by noting that the singlet algebra arises as the $\psi \to \infty$ limit of the corner algebra of Gaiotto-Rap{\v c}{\'a}k $Y_{1,1,0}[\psi]$, \cf\, \cite[Section 7.2]{GaiottoRapcak}, modulo decoupling a commutative $U(1)_0$ subalgebra. This decoupling process is analogous to that arising in the related construction of the full Feigin-Tipunin algebra $\CF\CT_k(\mathfrak{sl}(n))$ described in \cite[Section 6.2]{CDGG} -- the commutative subalgebra arising in the $\psi \to \infty$ limit should be understood as a deformation of the singlet algebra by a background flat connection for the corresponding $U(1)$ flavor symmetry. From this perspective, the Wilson lines we identified above are merely the ($\psi \to \infty$ limit of the) degenerate $W_n$ modules of \cite{GaiottoRapcak}.

\subsection{$B$ twist}
\label{sec:CSYMB}
Let us now move to deforming the $HT$-twisted theory given in Eq. \eqref{eq:twistedactionGW} to the topological $B$ twist. The $B$ twist deformation for the hypermultiplet fields is given by $\int \bY \pd \bX$, \cf\, the deformation in Section \ref{sec:SYMB}. In contrast with super Yang-Mills, the transformation of the connection shows that we should have $\delta_B \bA \sim -\tfrac{2\pi}{k}\bnu_\C$. In particular, we expect that the $B$-twisted action should take the form
\be
\label{eq:BtwistedactionGW}
S_B = S + \int \bY \pd \bX - \tfrac{2\pi}{k} \Tr(\bB \bnu_\C)\,.
\ee
We see that the action of $Q_B$ is given by%
\be
\begin{aligned}
	\label{eq:QBGW}
	Q_B \bA & = F'(\bA) - \tfrac{2\pi}{k}\bnu_\C \qquad & Q_B \bB & = \diff'_{\bA} \bB - \bnu + \tfrac{k}{2\pi}\pd \bA \\
	Q_B \bX & = \diff'_{\bA} \bX \qquad & Q_B \bPsi_{\bX} & = \diff'_{\bA} \bPsi_{\bX} - \pd \bY - \tfrac{2\pi}{k} \bY \bB\\
	Q_B \bY & = \diff'_{\bA} \bY \qquad & Q_B \bPsi_{\bY} & = \diff'_{\bA} \bPsi_{\bY} + \pd \bX - \tfrac{2\pi}{k}\bB \bX
\end{aligned}\,.
\ee
Crucially, we need the fundamental identity for this deformation to satisfy the classical master equation. For example, one finds%
\be
\label{eq:GWBnilcheck}
\begin{aligned}
	Q_B^2 \bX & = \big(F'(\bA) - \tfrac{2\pi}{k}\bnu_\C\big) \cdot \bX - \diff'_\bA(\diff'_\bA \bX)\\
	& = - \tfrac{2\pi}{k}\bnu_\C \bX = -\pd_\bY\big(\tfrac{\pi}{k}\Tr(\bnu_\C{}^2)\big)\\
\end{aligned}\,.
\ee
Just as in the $A$-twist, it is straightforward to check that this theory is perturbatively topological: the stress tensor is $\bT = - \bB \pd_z \bA + \bPsi_{\bX} \pd_z \bX + \bPsi_{\bY} \pd_z \bY$ and the homotopy trivialization is provided by $\bS = -\tfrac{\pi}{k}\Tr (\bB \iota_{\pd_z} \bB) - \bPsi_{\bX} \iota_{\pd_z} \bPsi_{\bY}$.

Before we move to the case with $\CN=4$ vector multiplets, it is worth noting that we can combine the Chern-Simons fields $\bA, \bB$ as
\be
\CA = \bA - \varepsilon \big(\tfrac{2\pi}{k} \bB\big)
\ee
in a way that is compatible with the hypermultiplet fields $\CX, \CY$. This dramatically simplifies the $B$-twisted action to%
\be
\label{eq:BtwistedactionGW2}
S_B = \int {\rm CS}(\CA) + \CY \diff_\CA \CX\,,
\ee
and the action of $Q_B$ to%
\be
\label{eq:QBSQCD2}
\begin{alignedat}{3}
	&& \hspace{-1 cm} Q_B \CA = F(\CA)  - \tfrac{2\pi}{k} \nu_\C \hspace{-1 cm} && \\
	Q_B \CX & = \diff_\CA \CX & & Q_B \CY & = \diff_\CA \CY\\
\end{alignedat}\,.
\ee

As expected, this is exactly the action for a Chern-Simons theory based on the Lie superalgebra $\widehat{\mathfrak{h}} = \mathfrak{h} \oplus \Pi(T^* U)$, \cf\, \cite{KapustinSaulina-CSRW}, with non-trivial brackets
\be
\label{eq:CSBliealg}
\begin{alignedat}{3}
	&& \hspace{-0.5 cm} [t_\alpha, t_\beta] = f^\gamma{}_{\alpha \beta} t_\gamma \hspace{-0.5 cm} && \\
	[t_\alpha, \theta_m] & = \theta_n (\sigma_\alpha)^n{}_m &&&  [t_\alpha, \otheta^n] &= -(\sigma_\alpha)^n{}_m \otheta^m\\
	&& \hspace{-1cm} \{\theta_m, \otheta^n\} = -\tfrac{2\pi}{k}\delta^{\alpha \beta}(\sigma_\alpha)^n{}_m t_\beta \hspace{-1.5cm} &&\\
\end{alignedat}\,,
\ee
where the $t_\alpha$ are a basis of $\mathfrak{h}$, $\sigma_\alpha$ are the representation matrices for the $\mathfrak{h}$ action on $U$, and $\delta^{\alpha \beta}$ are the matrix elements of the inverse to the pairing $\Tr$ in the basis $t_\alpha$. Phrased differently, we find that this theory can be described as an AKSZ theory of maps into the classifying space $B\widehat{H}$ for the Lie supergroup $\widehat{H}$ (whose underlying bosonic group is $H$ and whose Lie superalgebra is $\widehat{\mathfrak{h}}$)
\be
	\textrm{Solutions to EOM: } \quad \Maps(\R^3_{\dR}, B\widehat{H})\,.
\ee
From the perspective that this $B$-twisted Chern-Simons-matter theory is merely a supergroup Chern-Simons theory, the moment map for the action of the even subalgebra on the odd subspace (together with the bilinear form $\Tr$) determines the bracket of two odd elements, and the fundamental identity corresponds to the Jacobi identity for 3 odd elements.

Including $\CN=4$ vector multiplets in the above construction is somewhat more delicate than what arose in the ``$A$ twist." Consider an $\CN=4$ theory of vector multiplets for the gauge group $G$ and hypermultiplets valued in the representation $R = \bigoplus_i T^*(V_i \otimes U_i)$, where the $V_i$ are representations of $G$ and the $U_i$ are representations of the Higgs-branch flavor symmetry group $H$. We interpret the $B$ twist of this theory in the same way as above: the underlying bosonic group is $G \ltimes \mathfrak{g}^*$, where $\mathfrak{g}^*$ is viewed as an additive Lie group acted upon by $G$ in the coadjoint representation. The moment map for the action of this bosonic group on $R$ is given by $\wt{\mu}_\C = \mu_{\C} \oplus 0: R \to (\mathfrak{g} \oplus \mathfrak{g}^*)^* \cong \mathfrak{g}^* \oplus \mathfrak{g}$, \cf\, Eq. \eqref{eq:SYMBliealg}, and the Lie algebra pairing is simply the natural pairing between $\mathfrak{g}$ and $\mathfrak{g}^*$. The fundamental identity is trivially satisfied for these theories simply because $\mathfrak{g}^*$ acts trivially on $R$ (and itself).

Now assume that the action of the hypermultiplet flavor symmetry group $H$ (or perhaps a subgroup thereof) satisfies the fundamental identity on the Higgs branch. We will find that the right-hand side of Eq. \eqref{eq:fundid} is no longer zero but instead something times the complex moment map $\mu_{\C}$
\be
\label{eq:CSBfundid}
	\tfrac{2\pi}{k}\Tr(\nu_\C{}^2) - 2\mu_{\C} \xi_\C = 0\,.
\ee
This reduces to the usual fundamental identity on the Higgs branch $\mu_\C = 0$. The map $\xi_\C: R \to \mathfrak{g}$ has $U(1)_C$ charge 2, and gauge covariance of this equation implies that it is $H$-invariant and $G$-equivariant. We can therefore write it in components as $(\xi_\C)^a = Y_n (\rho^a)^n{}_m X^m$. If we think of $(\mu_{\C},\xi_\C)$ as the moment map for an action of $\mathfrak{g} \oplus \mathfrak{g}^*$ on $R$, it follows that Eq. \eqref{eq:CSBfundid} is simply another instance of the familiar fundamental identity, where we have taken the non-degenerate $G \times H$-invariant pairing on $\mathfrak{g} \oplus \mathfrak{g}^*$ to be (the negative of) the usual pairing of $\mathfrak{g}$ and $\mathfrak{g}^*$.

Since the moment map for $\mathfrak{g}$ is unchanged, its action on $R$ is unchanged. Similarly, due to the $G$-equivariance of $\xi_\C$, the brackets between $\mathfrak{g}$ and $\mathfrak{g}^*$ are unchanged. However, the non-trivial Poisson brackets between the components of $\xi_\C$ correspond to non-trivial brackets involving two elements of $\mathfrak{g}^*$. We will make the simplifying assumption that the Poisson bracket of two components of $\xi_\C$ is given by
\be
	\{(\xi_\C)^a, (\xi_\C)^b\}_{PB} = F^{abc} (\mu_{\C})_c \leftrightarrow [\rho^a, \rho^b] = F^{abc} \tau_c\,,
\ee
for $F^{abc}$ totally antisymmetric. These Poisson brackets correspond to the Lie algebra
\be
	[T_a, T_b] = f^c{}_{ab} T_c \qquad [T_a, S^b] = f^b{}_{ca}S^c \qquad [S^a, S^b] = F^{abc} T_c\,,
\ee
where $T_a$ are a basis of $\mathfrak{g}$ and $S^a$ are a dual basis of $\mathfrak{g}^*$ with respect to the natural pairing $\mathfrak{g} \otimes \mathfrak{g}^* \to \C$. By construction, the matrices $(\tau_a)^n{}_m, (\rho^a)^n{}_m$ represent an action of this Lie algebra on $\bigoplus_i V_i \otimes U_i$. The bracket of two odd elements then uses the pairings on $\mathfrak{h}$ and $T^*\mathfrak{g}$ and the representation matrices to give
\be
\label{eq:2oddcomm}
	\{\otheta^n, \theta_m\} = (\rho^a)^n{}_m T_a + (\tau_a)^n{}_m S^a - \tfrac{2\pi}{k}\delta^{\alpha \beta} (\sigma_{\alpha})^n{}_m t_\beta\,,
\ee
\cf\, Eq. \eqref{eq:CSBliealg}. We can then write a relatively straightforward combination of the twisted actions in Eq \eqref{eq:BtwistedactionYM2} and \eqref{eq:BtwistedactionGW2}:
\be
\label{eq:BtwistedactionCSYM}
	S_B = \int {\rm CS}(\CA_{\rm CS}) + \CB \big(F(\CA_{\rm YM}) + \tfrac{1}{3}\CB^2 + \xi_\C(\CX, \CY)\big) + \CY \diff_\CA \CX\,,
\ee
where $\CB^2 = \tfrac{1}{2}\CB_a \CB_b F^{abc} T_c$. The corresponding action of $Q_B$ is given by
\be
\label{eq:QBCSB}
\begin{alignedat}{3}
	&& \hspace{-1 cm} Q_B \CA_{\rm CS} = F(\CA_{\rm CS})  - \tfrac{2\pi}{k}\nu_\C \hspace{-1 cm} && \\
	Q_B \CA_{\rm YM} & = F(\CA_{\rm YM}) +\CB^2 + \xi_\C & & Q_B \CB & = \diff_{\CA_{\rm YM}} \CB + \mu_\C\\
	Q_B \CX & = \diff_\CA \CX + \CB \frac{\pd\xi_\C}{\pd \CY} & & Q_B \CY & = \diff_\CA \CY - \CB \frac{\pd\xi_\C}{\pd \CX}\\
\end{alignedat}\,.
\ee
This action solves the classical master equation as a result of Eq. \eqref{eq:CSBfundid}. For example, one finds that
\be
	Q_B^2 \CX = \frac{\pd}{\pd \CY} \bigg(-\tfrac{\pi}{k} \Tr(\nu_\C{}^2) + \mu_{\C} \xi_\C\bigg) = 0\,,
\ee
\cf\, Eq. \eqref{eq:GWBnilcheck}. The fact that this theory is perturbatively topological can be checked as above.

We are led to conclude that the resulting theory is simply another instance of Chern-Simons theory, now based on the above gauge Lie superalgebra $T^*\mathfrak{g} \oplus \mathfrak{h} \oplus \Pi R$. Unlike the previous cases, however, it isn't immediately obvious what the global form of the bosonic group should be, or even how seriously one should take this analogy beyond perturbation theory. For example, the full bosonic gauge group, whatever it is, should at least contain $G \times H$ as a subgroup. Note that the underlying physical QFT only contains monopole operators for this $G \times H$ subgroup and not the full bosonic group. For $\CN=4$ super Yang-Mills, one can safely ignore this problem because the full bosonic group is simply the total space of the cotangent bundle $T^*G \cong G \ltimes \mathfrak{g}^*$, where we view $\mathfrak{g}^*$ as an additive Lie group, whence the topology of the gauge group is essentially unchanged from the physical theory. We hope to return to this question in future work.

\subsubsection{Example: SQCD}
\label{sec:SQCD}

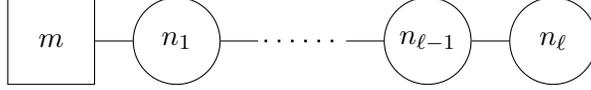
\begin{figure}[h!]
	\centering
	\begin{tikzpicture}[scale=1]
		\begin{scope}[auto, every node/.style={inner sep=1}, node distance=1cm]
			\node (v3) {};
			\node[right=of v3] (v4) {};
		\end{scope}
		\begin{scope}[auto, every node/.style={minimum size=3em,inner sep=1},node distance=0.5cm]
			\node[draw, circle, left=of v3] (v2) {$n_{1}$};
			\node[draw, left=of v2] (v1) {$m$};
			\node[draw, circle, right=of v4] (v5) {$n_{\ell-1}$};
			\node[draw, circle, right=of v5] (v6) {$n_\ell$};
		\end{scope}
		\draw(v1)--(v2);
		\draw(v2)--(v3);
		\draw[loosely dotted, line width = 1pt] (v3)--(v4);
		\draw(v4)--(v5);
		\draw(v5)--(v6);
	\end{tikzpicture}
	\caption{A linear quiver describing a family (parameterized by a positive integer $\ell$ and an $\ell$-vector of positive integers $\vec{n} = (n_1, ..., n_\ell)$) of 3d $\CN=4$ gauge theories that have an $SU(m)$ flavor symmetry that satisfies the fundamental identity on its Higgs branch. The special case $\ell = 1$ corresponds to SQCD.}
	\label{fig:linquiver2}
\end{figure}

To make the above more concrete, we consider the simple example of 3d $\CN=4$ SQCD with gauge group $U(n)$ with $m$ hypermultiplets in the fundamental representation. This example is the simplest instance of an infinite family of 3d $\CN=4$ quiver gauge theories whose $SU(m)$ flavor symmetry satisfies the fundamental identity. These quivers are of the form in Figure \ref{fig:linquiver2}; in particular, it includes the $T[SU(m)]$ quiver in Figure \ref{fig:linquiver}.

In the present SQCD example, the hypermultiplet scalar $X$ (resp. $Y$) can be thought of as a matrix $X \in \Hom(\C^m, \C^n)$ (resp. $Y \in \Hom(\C^n, \C^m)$). If we further identify $\mathfrak{gl}(n) \cong \mathfrak{gl}(n)^*$ with the trace pairing, the complex moment map can be written as $\mu_\C = X Y$. Similarly, the complex moment map for the action of the $\mathfrak{sl}(m)$ Higgs-branch flavor symmetry can be expressed as the matrix $\nu_\C = -Y X + \tfrac{1}{m}\Tr(Y X)\mathds{1}_m$, where $\mathds{1}_m$ is the $m \times m$ identity matrix. We will take the invariant paring $\Tr$ to be the trace taken in the fundamental representation $\C^m$. It follows that
\be
\label{eq:SQCDfundid}
\tfrac{2\pi}{k} \Tr(\nu_\C{}^2) = \tfrac{2\pi}{k} \big(\Tr(\mu_\C{}^2) - \tfrac{1}{m}\Tr(\mu_\C)^2\big)\,,
\ee
so the fundamental identity is indeed satisfied on the Higgs branch. We find
\be
	\xi_\C = \tfrac{\pi}{k}\big(\mu_{\C} - \tfrac{1}{m} \Tr(\mu_{\C}) \mathds{1}_n\big)\,.
\ee
In particular, the structure constants $F^{abc}$ from above are given by
\be
	F^{abc} = \tfrac{\pi^2}{k^2}\delta^{aa'} \delta^{bb'} f^c{}_{a'b'}\,,
\ee
where $\delta^{ab}$ are the matrix elements of the bilinear form $\Tr$ on $\mathfrak{gl}(n)^* \cong \mathfrak{gl}(n)$ in the basis $S^a$, and $f^a{}_{bc}$ are the structure constants of $\mathfrak{gl}(n)$ in the basis $T_a$.

We identify the Lie superalgebra underlying this SQCD example as follows. For $A \in \mathfrak{gl}(n)$ and $B \in \mathfrak{gl}(n)^* \cong \mathfrak{gl}(n)$, Poisson brackets with the above moment maps imply that their action on $X$ and $Y$ are given by
\be
\delta_{A,B} X = \bigg(A + \tfrac{\pi}{k}\big(B - \tfrac{1}{m} \Tr(B) \mathds{1}_n\big)\bigg)X \qquad \delta_{A,B} Y = -Y\bigg(A + \tfrac{\pi}{k}\big(B- \tfrac{1}{m} \Tr (B) \mathds{1}_n\big)\bigg)\,.
\ee
Similarly, the action of $M \in \mathfrak{sl}(m)$ is
\be
\delta_M X = -X M \qquad \delta_M Y = M Y\,.
\ee
We can neatly arrange these actions, together with Eq. \eqref{eq:2oddcomm}, as graded commutation relations in a matrix Lie superalgebra by arranging $A,B,M,X,Y$ into the following $(2n|m) \times (2n|m)$ super matrix
\be
(A,B; M; X,Y) \leftrightarrow \begin{pmatrix}
	A - \tfrac{\pi}{k}\big(B - \tfrac{1}{m} \Tr(B) \mathds{1}_n\big) & 0 & 0\\
	0 & A + \tfrac{\pi}{k} \big(B + \tfrac{1}{m} \Tr(B) \mathds{1}_n\big) & \tfrac{2\pi}{k} X\\
	0 & Y & M + \tfrac{2\pi}{m k} \Tr(B) \mathds{1}_m\\
\end{pmatrix}\,.
\ee
From this embedding, it is straightforward to check that the corresponding Lie superalgebra is $\mathfrak{gl}(n) \ltimes \mathfrak{sl}(n|m)$; the $\mathfrak{sl}(n|m)$ subalgebra is generated by $B' = \tfrac{2\pi}{k} B, X' = \tfrac{2\pi}{k}X,Y,M$ and the $\mathfrak{gl}(n)$ subalgebra is generated by $A' = A - \tfrac{\pi}{k}(B - \tfrac{1}{m} \Tr(B) \mathds{1}_n)$, acting on $B$ in the adjoint representation and $X'$ (resp. $Y$) in the fundamental (resp. anti-fundamental) representation:
\be
(A',B'; M; X,Y) \leftrightarrow \begin{pmatrix}
	A' & 0 & 0\\
	0 & A' + B' & X'\\
	0 & Y & M + \tfrac{1}{m} \Tr(B') \mathds{1}_m\\
\end{pmatrix}\,.
\ee
\subsubsection{Boundary algebra for the $B$ twist}
\label{sec:bdyCSYMB}
We now describe a boundary algebra for the $B$-twisted Chern-Simons-matter theories described introduced above. When we do not include $\CN=4$ vector multiplets, these $B$-twisted theories were studied in depth in \cite{KapustinSaulina-CSRW} and, as we saw above, the twisted action agrees with that of Chern-Simons theory for a supergroup $\widehat{H}$ at level $k$.

For these simple $B$-twisted Chern-Simons-matter theories, the boundary conditions we impose are the following.
\begin{itemize}
	\item Dirichlet boundary conditions for the $\CN = 2$ vector multiplets. ($\bA|_{\pd} = 0$)
	\item Dirichlet boundary conditions for the hypermultiplets. ($\bX|_{\pd}, \bY|_{\pd} = 0$)
\end{itemize}
The choice of Dirichlet boundary conditions for the gauge fields will result in boundary monopole operators. We will leave the full, non-perturbative analysis of boundary monopoles to future work and only consider the perturbative part of the boundary algebra.

The perturbative analysis is nearly identical to the previous cases. The only fields that survive cohomology at the boundary are the bottom components $B, \psi_X, \psi_Y$, and $Q_B$ acts trivially on them after imposing the boundary conditions. The OPEs of these operators (after rescaling%
\footnote{This rescaling sends the holomorphic symplectic form $\Omega$ to $\tfrac{k}{2\pi}\Omega$. Thus, the non-degenerate pairing on $\widehat{\mathfrak{h}}$ is given by $k(\Tr \oplus \Omega)$.})
take the following form:
\begin{gather}
	B_\alpha(z) B_\beta(w) \sim  \frac{(k - h_H + T_U)\delta_{\alpha \beta}}{(z-w)^2} + \frac{f^\gamma{}_{\alpha \beta}}{z-w} B_\gamma(w) \nonumber\\
	\qquad B_\alpha(z) \psi_Y{}^n(w) \sim -\frac{(\sigma_\alpha)^n{}_m}{z-w}\psi_Y{}^m(w) \quad \quad B_\alpha(z) \psi_X{}_n(w) \sim \frac{(\sigma_\alpha)^m{}_n}{z-w}\psi_X{}_m(w) \label{eq:opeGWB}\\
	\psi_Y{}^n(z) \psi_X{}_m(w) \sim \frac{(k- C_2(U)) \delta^n{}_m}{(z-w)^2} - \frac{(\sigma_\alpha)^n{}_m }{z-w}   \delta^{\alpha \beta} B_\beta(w)\nonumber
\end{gather}
where $\delta_{\alpha \beta}$ and $\delta^{\alpha \beta}$ denote the matrix elements of the bilinear form $\Tr$ and its inverse, and $C_2(U)\delta^n{}_m := \delta^{\alpha \beta} (\sigma_\alpha \sigma_\beta)^n{}_m$ is the quadratic Casimir for the $\mathfrak{h}$-representation $U$ (with respect to the bilinear form $\Tr$). We note that contributions to the second-order pole in the $BB$ OPE come from the bulk Chern-Simons term and the 1-loop diagrams coming from the trivalent gauge vertices. Similarly, the second-order pole in the $\psi_X \psi_Y$ OPE receives contributions from the $B$ twist deformation $\bY\pd \bX$ as well as the 1-loop diagrams that connect a $\Tr(\bB \bnu_\C)$ vertex to either $\bPsi_{\bX} \bA \bX$ or $\bY \bA \bPsi_{\bY}$.

For any representation $T^*U$ such that Eq. \eqref{eq:CSBliealg} defines a simple Lie superalgebra, it follows that the quadratic Casimir $C_2(U)$ satisfies the relation%
\footnote{This relation can be determined by considering the value of the Casimir operator $C_2 = \delta^{\alpha \beta}t_\alpha t_\beta - \theta_n \otheta^n + \otheta^n \theta_n$ on the even and odd subspaces, \cf\, \cite[Section 2]{GOWsugawara}. The value of $C_2$ on each simple factor $\mathfrak{h}_i$ of the even subalgebra is $2(h_i-T_{U,i})$, where $h_i$ is the dual Coexeter number of $\mathfrak{h}_i$ and $T_{U,i}$ is the quadratic index of $U$ viewed as a $\mathfrak{h}_i$ representation. Similarly, it's value on each irreducible $\mathfrak{h}$ representation $R_n \subseteq U$ is $2 C^{\rm tot}_2(R_n)$. Since $\widehat{\mathfrak{h}}$ is simple, $C_2$ must take a single value in the adjoint representation, whence $h_i-T_{U,i} = C^{\rm tot}_2(R_n)$ for every $i$ and every $n$.} %
\be
	C_2(U) = h_H - T_U\,
\ee
For example, if we consider $\mathfrak{h} = \mathfrak{sl}(N) \oplus \mathfrak{sl}(M) \oplus \mathfrak{u}(1)$ with $U = \Hom(\C^M, \C^N)$, corresponding to $\mathfrak{sl}(N|M)$ for $N \neq M$, one finds $C_2(U) = (\tfrac{N^2-1}{N}) - (\tfrac{M^2-1}{M}) + (\tfrac{1}{N} - \tfrac{1}{M}) = N-M$, which agrees with $(h_H - T_U)\delta^{\alpha \beta}$ for each of the bosonic factors.%
\footnote{The $\mathfrak{u}(1)$ factor is generated by the diagonal matrix $t_0 = \diag(\tfrac{1}{N}, ..., \tfrac{1}{N}, \tfrac{1}{M}, ..., \tfrac{1}{M})$, and so $U$ is simply $NM$ copies of the charge $\tfrac{1}{N}-\tfrac{1}{M}$ representation. One finds $t_0$ is orthogonal to $\mathfrak{sl}(N) \oplus \mathfrak{sl}(M)$ and $\delta_{00} = \tfrac{1}{N}-\tfrac{1}{M}$, whence the index $T_{U, \mathfrak{u}(1)}$ is $NM(\tfrac{1}{N} - \tfrac{1}{M}) = M-N$.} %
Thus, putting the above together, we find a $\widehat{\mathfrak{h}}$ current algebra on the boundary with critically shifted level $k$. This perturbative current algebra admits a Sugawara stress tensor
\be
	T(z) = \tfrac{1}{2k}\big(\delta^{\alpha \beta}\norm{B_\alpha B_\beta} - \norm{\psi_X{}_n \psi_Y{}^n} + \norm{\psi_Y{}^n \psi_X{}_n}\big)
\ee
of central charge $c = \tfrac{k-h_H+T_U}{k} (\dim \mathfrak{h} - 2\dim U)$, so long as $k \neq 0$. For $\mathfrak{sl}(N|M)$, we find a central charge $c = \tfrac{k - N + M}{k}((N-M)^2+1)$.

When we include $\CN=4$ vector multiplets, we take an admixture of the boundary condition above and the one used in Section \ref{sec:SYMB}:
\begin{itemize}
	\item Dirichlet boundary conditions for the vector multiplets. ($\bA|_{\pd} = 0$)
	\item Neumann boundary conditions for the adjoint chiral multiplet. ($\bLambda|_{\pd} = 0$)
	\item Dirichlet boundary conditions for the hypermultiplets. ($\bX|_{\pd}, \bY|_{\pd} = 0$)
\end{itemize}
Again, we will only consider a perturbative analysis and leave an understanding of the full, non-perturbative algebra to future work.

Just as above, the perturbative boundary algebra is generated by the lowest components $B_{\rm CS}, B_{\rm YM}, \phi, \psi_X, \psi_Y$ and the perturbative OPEs are relatively straightforward to determine: we find a current algebra based on the Lie superalgebra $T^*\mathfrak{g} \oplus \mathfrak{h} \oplus \Pi R$. The first-order poles are determined by the brackets in $T^*\mathfrak{g} \oplus \mathfrak{h} \oplus \Pi R$, and the coefficients of the second-order poles are determined by the bulk Chern-Simons term, the $B$ twist deformation $\bY \pd \bX - \bLambda \pd \bA$, and various 1-loop diagrams involving the various trivalent vertices.

We can compute the half-index for this boundary condition using the methods of \cite{DimofteGaiottoPaquette}. In the absence of monopole operators, the half-index reads
\be
\label{eq:halfind0CSYMB}
	I\!\!I_0(q,s,t) = \frac{\prod\limits_i \prod\limits_{\lambda \in {\rm wt}_{G}(V_i)}\prod\limits_{\omega \in {\rm wt}_{H}(U_i)}(q s^{\lambda} t^{\omega}; q)_\infty (q s^{-\lambda} t^{-\omega}; q)_\infty}{\prod\limits_{\alpha \in {\rm wt}_G(\mathfrak{g})} (q s^{\alpha}; q)_\infty (q s^{-\alpha}; q)_\infty \prod\limits_{\beta \in {\rm wt}_H(\mathfrak{h})} (q t^{\beta}; q)_\infty}\,,
\ee
where ${\rm wt}_G(R_G)$ denotes the weights of the $G$ representation $R_G$, and similarly for ${\rm wt}_H(R_H)$. The variables $s, t$ are fugacities for the gauge tori $T_G \times T_H$. This exactly matches the character for the vacuum module of the perturbative current algebra; the numerator counts the positive modes of $\psi_Y, \psi_X$ and the denominator counts the positive modes of $B_{\rm YM}, \phi, B_{\rm CS}$. Boundary monopole contributions are counted by a sum over abelian monopole charges, \ie\, the cocharacter lattice of a maximal torus $T_G \times T_H \subset G \times H$:
\be
\label{eq:halfindCSYMB}
	I\!\!I(q,s,t) = \sum\limits_{\mathfrak{m} \in {\rm cochar}(T_G)}\sum\limits_{\mathfrak{n} \in {\rm cochar}(T_H)} I\!\!I_0(q,q^{\mathfrak{m}} s,q^{\mathfrak{n}} t) q^{\tfrac{1}{2}(k_G \Tr_G \mathfrak{m}^2 + k_H \Tr_H \mathfrak{n}^2)} s^{k_G \mathfrak{m}} t^{k_H \mathfrak{n}}\,,
\ee
where $k_G, k_H$ are the effective Chern-Simons levels for $G,H$, \ie\, the level of the boundary $\mathfrak{g}$ and $\mathfrak{h}$ currents. Note that there is no sum over monopole charges for the full bosonic current algebra; one should only include monopole operators for the physical gauge group $G \times H$.

Instead of writing down the general form of the perturbative current algebra, we focus on the example discussed above for $n = 1$ and $m = 2$, \ie\, we will consider a Chern-Simons gauging of the $SU(2)$ flavor symmetry of $\CN=4$ SQED with two flavors.

\subsubsection{Example: SQED revisited}
\label{sec:SQEDrev}
The example of SQED with two hypermultiplets is particularly interesting as it flows to $T[SU(2)]$ in the IR. In particular, this analysis should realize the $B$ twist counterpart of the $A$ twist analysis of \cite{CDGG}. Moreover, we expect that the corresponding $B$-twisted theory agrees with the $SU(2)$ Chern-Simons-Rozansky-Witten theory (with target $T^*\P^1$) considered in \cite{KapustinSaulina-CSRW}. Kapustin and Saulina made concrete proposals for the algebra of local operators and the category of line operators in this TQFT, and these can be compared to the self-extensions and modules of this boundary algebra (once it has been extended by monopole operators for $U(1)$ and $SU(2)$).

SQED has two charge 1 hypermultiplets $X^n, Y_n$, $n = 1,2$, with complex moment map $\mu_\C = Y_n X^n$. Similarly, the complex moment map for the $SU(2)$ flavor symmetry can be identified with a $2 \times 2$ matrix with components $(\nu_\C)^n{}_m = -Y_m X^n + \tfrac{1}{2} Y_{l} X^{l} \delta^n{}_m$. From Section \ref{sec:SQCD}, we find that the fundamental identity is satisfied
\be
	\tfrac{2\pi}{k} \Tr(\nu_\C{}^2) = \tfrac{\pi}{k} Y_m X^m Y_n X^n = 2 \xi_\C \mu_{\C}\,
\ee
where $\xi_\C = \tfrac{\pi}{2 k} Y_n X^n$. We interpret $\mu_{\C}, \xi_\C, \nu_\C$ as moment maps for a deformed action of $\C^2 \oplus \mathfrak{sl}(2)$, so that $X$ transforms with weight $1, \tfrac{\pi}{2k}$ under $T^*\C \cong \C^2$ and in the fundamental representation of $\mathfrak{sl}(2)$. These moment maps encode a Lie superalgebra with even generators $T,S$ (for $T^*\C$) and $t_\alpha$ (for $\mathfrak{sl}(2)$) and odd generators $\theta_n, \otheta^n$ with non-trivial brackets (after rescaling)
\be
\begin{split}
	[t_\alpha, t_\beta] = f^\gamma{}_{\alpha \beta} t_\gamma \hspace{3cm}\\
	[t_\alpha, \theta_n] = \theta_m (\sigma_{\alpha})^m{}_n \qquad [T, \theta_n] = \theta_n \qquad [S, \theta_n] = \tfrac{1}{4} \theta_n \hspace{0.5cm} \\
	[t_\alpha, \otheta^n] = -(\sigma_{\alpha})^n{}_m \otheta^m \qquad [T, \otheta^n] = -\otheta^n \qquad [S, \otheta^n] = -\tfrac{1}{4} \otheta^n\\
	\{\otheta^n, \theta_m\} = (\tfrac{1}{4} T + S)\delta^n{}_m - \delta^{\alpha \beta}(\sigma_\alpha)^n{}_m t_\beta \hspace{2cm}\\
\end{split}\,.
\ee
It is straightforward see that this Lie superalgebra is $\mathfrak{gl}(1|2)$.

Now we move on to the perturbative algebra of boundary local operators. We start with OPEs involving the $SU(2)$ current $B_{\alpha}$. The only non-trivial OPEs it can have are with itself and with $\psi_X, \psi_Y$. In particular, they are unchanged from those in Eq. \eqref{eq:opeGWB}:
\be
\label{eq:opeCSTSU2B1}
\begin{split}
	B_\alpha(z) B_\beta(w) \sim  \frac{(k-1)\delta_{\alpha \beta}}{(z-w)^2} + \frac{f^\gamma{}_{\alpha \beta}}{z-w} B_\gamma(w) \hspace{3cm}\\
	B_\alpha(z) \psi_Y{}^n(w) \sim -\frac{(\sigma_\alpha)^n{}_m}{z-w}\psi_Y{}^m(w) \qquad B_\alpha(z) \psi_X{}_n(w) \sim \frac{(\sigma_\alpha)^m{}_n}{z-w}\psi_X{}_m(w)\\
\end{split}\,,
\ee
where we used $h_{SL(2)} = 2$ and $T_{\rm fund} = 1$ for the fundamental representation of $SU(2)$.

Now consider OPEs involving the $U(1)$ current $B$. It can have non-trivial OPEs with itself and $\phi$, as well as $\psi_X$ and $\psi_Y$. The OPEs with itself and with $\psi_X, \psi_Y$ are identical to those appearing in Eq. \eqref{eq:opeSYMB}:
\be
\label{eq:opeCSTSU2B2}
\begin{split}
	B(z) B(w) \sim  \frac{2}{(z-w)^2}\hspace{4cm}\\
	B(z) \psi_Y{}^n(w) \sim -\frac{1}{z-w}\psi_Y{}^n(w) \quad \quad B(z) \psi_X{}_n(w) \sim \frac{1}{z-w}\psi_X{}_n(w)\\
\end{split}\,.
\ee
More interestingly, the OPE between $B$ and $\phi$ is deformed from Eq. \eqref{eq:opeSYMB} by the two 1-loop diagrams with one $\bX \bPsi_{\bX}$ edge and one $\bY \bPsi_{\bY}$ edge. Each diagram contributes $\tfrac{1}{2}(\rho)^n{}_m (\tau)^m{}_n = \tfrac{1}{4}$ to the second order pole (after rescaling $\phi$ appropriately)%
\footnote{The appearance of $-k$, as opposed to simply $k$, comes from the fact that the have a $-1$ Chern-Simons level coming from the $B$-twist deformation, \cf\, Eq. \eqref{eq:opeSYMB}.}:
\be
\label{eq:opeCSTSU2B3}
	B(z) \phi(w) \sim \frac{(\tfrac{1}{2} - k)}{(z-w)^2}\,.
\ee
Similarly, we find that there is a non-trivial OPE between $\phi$ and itself coming from the 1-loop diagrams with two $\bX \bPsi_{\bX}$ edges or two $\bY \bPsi_{\bY}$ edges; each contributes $\tfrac{1}{2} (\rho)^n{}_m (\rho)^m{}_n = \tfrac{1}{2}$ yielding the OPE
\be
\label{eq:opeCSTSU2B4}
	\phi(z) \phi(w) \sim \frac{\tfrac{1}{8}}{(z-w)^2}\,.
\ee
It also has a non-trivial OPE with the fermionic currents due to the deformation term $-(\bPsi_{\bX} \bLambda \bX + \bY \bLambda \bPsi_{\bY})$:
\be
	\phi(z) \psi_Y{}^n(w) \sim -\frac{\tfrac{1}{4}}{z-w}\psi_Y{}^n(w) \quad \quad \phi(z) \psi_X{}_n(w) \sim \frac{\tfrac{1}{4}}{z-w}\psi_X{}_n(w)\,.
\ee

Finally, we consider the OPE of $\psi_Y$ and $\psi_X$. There is a first order pole coming from the $\bY \bPhi \bX$, $\bY \bB_{\rm YM} \bX$, and $\bY \bB_{\rm CS} \bX$ vertices. A priori, the second order pole receives contributions from the $\bY \pd \bX$ vertex and the two 1-loop diagrams connecting the $\Tr (\bB_{\rm CS} \bnu_{\C})$ vertex and the Chern-Simons gauge vertices as well as the four 1-loop diagrams coming from connecting the $\bY \bB_{\rm YM} \bX$ (resp. $\bY \bPhi \bX$) vertex to either $\bPsi_{\bX} \bA_{\rm YM} \bX$ or $\bY \bA_{\rm YM} \bPsi_{\bY}$ (resp. $\bY \bLambda \bPsi_{\bY}$ or $\bPsi_{\bX} \bLambda \bX$). Thankfully, the latter 1-loop diagrams cancel in pairs. For example, the loop with one $\bPsi_{\bX} \bX$ edge and one $\bB_{\rm YM} \bA_{\rm YM}$ edge is canceled by the loop with one $\bPsi_{\bY} \bY$ edge and one $\bPhi \bLambda$ edge. We conclude that the OPE is given by
\be
\label{eq:opeCSTSU2B5}
	\psi_Y{}^n(z) \psi_X{}_m(w) \sim \frac{(k-1)\delta^n{}_m}{(z-w)^2} + \frac{1}{z-w}  \bigg(\delta^n{}_m\big(\tfrac{1}{4} B(w) + \phi(w)\big) - (\sigma_\alpha)^n{}_m \delta^{\alpha \beta} B_\beta(w)\big) \bigg)\,.
\ee

It is straightforward to see that the bosonic currents $B_\alpha$ and $\wt{B} := 2\phi + \tfrac{1}{2} B$ together with the fermionic currents $\psi_X$ and $\psi_Y$ generate an $\mathfrak{sl}(2|1)$ current algebra at critically shifted level $k$. Similarly, $B' = 2\phi - \tfrac{1}{2}B$ is a abelian current at level $2k$ and entirely decouples from the perturbative algebra.

As mentioned several times above, non-perturbative corrections extend this current algebra by boundary monopole operators for the bulk gauge group. Note, however, one should not extend by monopole operators/integrable representations for the full bosonic subalgebra: one should only extend by boundary monopoles for the abelian current $B = \wt{B}-B'$ and the $\mathfrak{su}(2)$ currents $B_\alpha$. This is supported by the computing the corresponding half-index:
\be
\label{eq:halfindSQED}
	I\!\!I(q,s,t) = \sum\limits_{\mathfrak{m} \in \Z}\sum\limits_{\mathfrak{n} \in \Z} I\!\!I_0(q,q^{\mathfrak{m}} s,q^{\mathfrak{n}} t) q^{\mathfrak{m}^2 + (k-1) \mathfrak{n}^2} s^{2\mathfrak{m}} t^{(k-1)\mathfrak{n}}\,,
\ee
where the perturbative half-index is given by
\be
\label{eq:halfind0SQED}
	I\!\!I_0(q,s,t) = \frac{(q s t; q)_\infty (q s^{-1} t; q)_\infty (q s^{-1} t^{-1}; q)_\infty (q s t^{-1}; q)_\infty}{(q; q)_\infty{}^3 (q t^{2}; q)_\infty (q t^{-2}; q)_\infty}\,.
\ee
The numerator counts the positive modes of $\psi_Y{}^n,\psi_X{}_n$ for $n = 1,2$, and the denominator counts the modes of $B,\phi, B_\alpha$ for $\alpha = 1,2,3$.

\section{Future directions}
There are many interesting future directions and applications of this work, some of which are the following.

\subsection{State spaces, local operators, and descent}

The twisted formalism we work with is compatible with spacetimes that locally look like $\R_t \times \C_{z,\oz}$. By analyzing the equations of motion on a Riemann surface $\Sigma$ for the theories described in this paper, and performing geometric quantization of the resulting phase space, one has direct access to the BPS state spaces of our topological Chern-Simons-matter theories. This was done explicitly in \cite{twistedN=4} for the $A$ and $B$ twist of standard $\CN=4$ gauge theories, and it should be possible to extend those results to the theories studied in this paper: an abstract definition and qualitative aspects of these state spaces were discussed in \cite{CDGG} for the $A$ twist of $T[SU(n)]/SU(n)_k$. Alternatively, one should be able to access these same BPS state spaces by considering the (derived) spaces of conformal blocks of the boundary VOAs we propose.

Both of these abstract descriptions can be quite difficult to compute in general. Nonetheless, it should be possible to understand quantitative aspects of these BPS state spaces using more common techniques available to study supersymmetric gauge theories. For example, the twisted indices \cite{BZ-index, BZ-Riemann, CK-comments} of these theories can be computed using a variety of methods. In particular, the Bethe root analysis of \cite{NS-Bethe, NS-int} contains information about the category of line operators and symmetries; the recent paper \cite{CDGG} uses these techniques to great effect.

A particularly interesting application of an explicit description of the BPS state spaces would be to better understand the full, non-perturbative algebra of local operators in the theories we study. The recent paper \cite{Zeng} performed the geometric quantization mentioned above (for $\Sigma = S^2$) in the context $HT$-twisted $\CN=2$ abelian gauge theories, and was able to explicitly compare the resulting algebras to computations performed using the boundary algebras of \cite{CostelloDimofteGaiotto-boundary}. Most of the examples studied in this paper are outside the scope of \cite{Zeng}, being non-abelian gauge theories, but it would be interesting to see how, \eg\,, the $HT$-twist algebra of $U(1)_1$ with a hypermultiplet gets deformed to the ``$A$-twist."

Once one has a handle on local operators, one is naturally drawn to understand the secondary products arising from topological descent \cite{descent}. The work \cite{twistedN=4} discusses how this secondary product is inherited from the secondary product arising from descent in the $HT$ twist \cite{CostelloDimofteGaiotto-boundary}, and it should be possible to extend those results to understanding secondary products of local operators in the theories studied in this paper.

\subsection{Line operators and boundary conditions}

The twisted formalism can be used to describe aspects of extended operators, including line operators extended in the topological direction, boundary conditions/interfaces on a transverse plane, and intersections between different dimensional operators. 

As described for more standard $\CN=4$ theories in \cite{twistedN=4}, it is possible to access aspects of the category of line operators by geometric quantization of the ($1$-shifted) symplectic space of solutions to the equations of motion on the (formal) punctured disk $\D^\times$. Not much is known about the categorical properties of BPS line operators compatible with the twists we consider in this paper, but the twisted formalism presented in this paper could provide a useful avenue for their study. The main exception to this is the $B$-twisted Gaiotto-Witten theory studied by \cite{KapustinSaulina-CSRW}, where it was argued to be related to a deformed category of coherent sheaves. It would be interesting to derive this result directly from field theory using the results of this paper.

Boundary conditions are central objects in this paper, but there are a myriad of other choices that should provide complementary insight into other aspects of the bulk theory. We mainly focus on boundary conditions that preserve some amount of chiral supersymmetry, namely $(0,2)$ or $(0,4)$, and admit chiral algebras of boundary local operators. Even at this level, there could be other interesting choices we do not consider, \eg\, Dirichlet boundary conditions for the Chern-Simons gauge fields in the $A$-twisted Gaiotto-Witten theories. Additionally, there may be choices of boundary conditions that are topological in nature and admit ordinary associative algebras of local operators; such topological boundary conditions are particularly important for more standard $\CN=4$ theories \cite{BDGH, BDGHK, HKW, DGGH, GK}.

\subsection{Generalized Gaiotto-Witten theories}
Another class of theories that can easily be approached with the techniques of this paper are the generalized Gaiotto-Witten theories of \cite{HLLLP1, HLLLP2}. This generalization allows for a far richer collection of $\CN \geq 4$ theories at the expense of introducing twisted hypermultiplets and superpotentials coupling the two types of hypermultiplets. These can be yet further generalized to non-linear hypermultiplet and/or twisted hypermultiplet targets, possibly gauged with $\CN=4$ vector multiplets, so long as the fundamental identity is suitably satisfied. The mechanism that allows for these highly supersymmetric Chern-Simons-matter theories was realized as a consequence of the representation theory of metric Lie algebras in \cite{dMFOME} and its M-theory origin was discussed in \cite{N=34}. Some particularly noteworthy specimens in this collection of Chern-Simons-matter theories are the theories of Aharony-Bergman-Jaffries-Maldacena (ABJM) \cite{ABJM}, Aharony-Bergman-Jaffries (ABJ) \cite{ABJ}, and Bagger-Lambert-Gustovsson (BLG) \cite{BL1,BL2,Gustavsson} modeling the low energy dynamics of multiple M2 branes.

Boundary conditions for the highly supersymmetric BLG and ABJM theories are particularly interesting. For example, the work \cite{BPSTboundary} proposed an interpretation of several BPS boundary conditions for the BLG and ABJM theories as describing the low energy dynamics of M2 branes ending on various other M-theory objects. Closer in spirit to the present analysis is \cite{OSboundary}, where it was argued that the topologically twisted BLG and ABJM theories admit certain supergroup WZW models on their boundary and propose that they describe the low energy dynamics of multiple M2 branes ending on the intersection of two non-parallel M5 branes. We hope to analyze these boundary conditions and perhaps relate them to twisted holographic studies of, \eg\,, \cite{CGholography, GaiottoRapcak2, OhZhou, AbajianGaiotto} in the future.

\acknowledgments
We would like to thank Tudor Dimofte for his support during the preparation of this paper and his suggestion for investigating this problem. We would also like thank Thomas Creutzig, Justin Hilburn, Brian Williams, and Keyou Zeng for useful conversations during the development of this project. N.G. acknowledges support from the University of Washington and previous support from T. Dimofte's NSF CAREER grant DMS 1753077.


\bibliography{topCSM}
\bibliographystyle{JHEP_TD}
	
\end{document}